\documentstyle[12pt,epsf,twoside]{article}

\newcommand{\sss}{\scriptscriptstyle}

\newcommand{\susy}{{\sc SuSy}}

\newcommand{\tgb}{\tan\beta} 
\newcommand{\epem}{e^{\sss +}e^{\sss -}}   
\newcommand{\Z}{Z^{\sss 0}}   
\newcommand{\Wpm}{W^{\sss \pm}}
\newcommand{\h}{h^{\sss 0}} 
\newcommand{\A}{A^{\sss 0}}

\newcommand{\phino}{\tilde{\gamma}} 
\newcommand{\Zino}{\tilde{Z}} 
\newcommand{\sthw}{\sin\theta_{\sss W}}
\newcommand{\cthw}{\cos\theta_{\sss W}}
\newcommand{\tgw}{\tan\theta_{\sss W}} 
\newcommand{\tgwq}{\tan^2\!\theta_{\sss W}} 
\newcommand{\sthwq}{\sin^2\!\theta_{\sss W}}
\newcommand{\cthwq}{\cos^2\!\theta_{\sss W}}
\newcommand{\sinb}{\sin\beta} 
\newcommand{\cosb}{\cos\beta} 
\newcommand{\sindb}{\sin 2\beta}
\newcommand{\cosdb}{\cos 2\beta}

\def\nrad{{$\tilde\chi_2^0 \to \tilde\chi_1^0 \gamma$}}

\def\n#1{\tilde{\chi}^{\sss 0}_{#1}}

\def\hino#1{\tilde{H}_{#1}}
\def\cp#1{\tilde{\chi}^{\sss +}_{#1}}
\def\cm#1{\tilde{\chi}^{\sss -}_{#1}} 
\def\cpm#1{\tilde{\chi}^{\sss \pm}_{#1}} 
\def\cmp#1{\tilde{\chi}^{\sss \mp}_{#1}} 

\def\mn#1{m_{\tilde{\chi}^0_{#1}}} 
\def\mc#1{m_{\tilde{\chi}^{\pm}_{#1}}} 

\def\miss#1{\not\!\!#1}

\newcommand{\eegg}{\mbox{$\epem\gamma\gamma + \miss{E_T}$}}

\def\ltap{\;\raisebox{-.5ex}{\rlap{$\sim$}} \raisebox{.5ex}{$<$}\;}
\def\gtap{\;\raisebox{-.5ex}{\rlap{$\sim$}} \raisebox{.5ex}{$>$}\;}

\def\beq{\begin{equation}}
\def\eeq{\end{equation}}
\def\bea{\begin{eqnarray}}
\def\eea{\end{eqnarray}}
\def\beast{\begin{eqnarray*}}
\def\eeast{\end{eqnarray*}}

\begin{document}
\pagestyle{myheadings} 
\markboth
{\it S.~Ambrosanio and B.~Mele / Supersymmetric Scenarios...}
{\it S.~Ambrosanio and B.~Mele / Supersymmetric Scenarios...}

\begin{titlepage}
\thispagestyle{empty} 
\begin{flushright} 
ROME1-1148/96  \\ 
July 1996      \\ 
hep-ph/9609212 \\ 
Revised: May 1997
\end{flushright} 

\vspace{1.0cm} 
					
\begin{center}
{\Large \bf Supersymmetric Scenarios with Dominant} \\ 
\vspace{0.2cm} 
{\Large \bf Radiative Neutralino Decay}~$^{\dagger}$ \\ 
\end{center}
~\\
\begin{center}
{\large 
Sandro~Ambrosanio~$^a$ and Barbara~Mele~$^b$
} \\  

~\\ 

\noindent
$^a$ {\it Randall Physics Laboratory}, University of Michigan, 
Ann Arbor, Michigan 48109-1120 \\
$^b$ INFN, Sezione di Roma 1 and University {\it La Sapienza}, 
Rome, Italy \\

\vspace{2.0cm}  

{\bf Abstract} \\
\end{center}
{\small 
The radiative decay of the next-to-lightest neutralino
into a lightest neutralino and a photon is analyzed
in the minimal supersymmetric standard model. 
We find that significant regions of the supersymmetric parameter 
space with large radiative branching ratios (up to about 100\%)  
do exist. The radiative channel turns out to be enhanced when the 
neutralino tree-level decays are suppressed either ``kinematically'' 
or ``dynamically''. In general, in the regions allowed by data 
from CERN LEP and not characterized by asymptotic values of the 
supersymmetric parameters, the radiative enhancement requires 
$\tgb \simeq 1$ and/or $M_1 \simeq M_2$, and negative values of $\mu$. 
We present typical specific scenarios where these {\it necessary} 
conditions are satisfied, relaxing the usual relation 
$M_1=\frac{5}{3}\tgwq M_2$. The influence of varying the top-squark 
masses and mixing angle when the 
radiative decay is enhanced is also considered. Some phenomenological 
consequences of the above picture are discussed. \protect\\ ~\protect\\  
{\bf PACS number(s)}: 14.80.Ly, 12.60.Jv
}

\vspace*{\fill}  

\noindent
\underline{e-mail}: \\
{\sl ambros@umich.edu, mele@roma1.infn.it} 
~\\ 
\parbox{0.4\textwidth}{\hrule\hfill} \\ 
{\small 
$^{\dagger}$\ Published (revisions to appear) in Physical Review D.   
}  

\end{titlepage}

\setcounter{page}{0} 
\thispagestyle{empty} 
~\\ 
\newpage 
\thispagestyle{plain} 

\section{Introduction}
\noindent 
Neutralinos ($\tilde\chi_i^0, \; i=1, \ldots 4; \; \mn{1} \le \ldots
\le \mn{4}$) are among the lightest supersymmetric (\susy) partners 
predicted in the  Minimal Supersymmetric Standard Model (MSSM) 
\cite{Haber-Kane}.\footnote{In this paper, by MSSM we refer to the
supersymmetric extension of the Standard Model with minimal gauge group
and particle content. No additional assumptions (such as unification
assumptions at a large scale) are understood, unless explicitly stated.}
In particular, the $\n{1}$ is usually the lightest \susy\ particle (LSP).
Hence, for conserved $R$-parity, $\n{1}$ is always present among the decay 
products of any superpartner, giving rise to large amounts of missing energy
and momentum in the final states corresponding to pair production of 
\susy\ particles in $e^+e^-/p\bar{p}/ep$ collisions.
The heavier neutralinos have in general a rather complicated decay pattern 
towards the LSP, with possible intermediate steps involving other
neutralinos and/or charginos as well as two-body decays with on-shell 
$\Z/\Wpm$, Higgses or sfermions in the final states \cite{Gunion-Haber}.  
Being the lightest visible neutralino, the next-to-lightest neutralino  
$\n{2}$ is of particular practical interest. It would be among the first 
\susy\ partners to be produced at the CERN $e^+e^-$ collider LEP2 and at 
the Fermilab TeVatron \cite{Ambr-Mele-1,Baer}. 

The dominant $\n{2}$ decay channels are, in general, tree-level decays 
into a lightest neutralino and two standard fermions through either a 
(possibly on-shell) $\Z$ or sfermion exchange.
Accordingly, the $\n{2}$ decays into a $\n{1}$ plus a $\ell^+\ell^-$, 
$\nu_\ell\bar{\nu}_\ell$ or $q\bar{q}$ pair. When $\n{2}$ is heavier 
than the lightest chargino, also cascade decays through a $\cpm{1}$ can 
become relevant. Hence, one can have, as final states,  
$\ell^+\ell^{\prime -}\nu_\ell\bar{\nu}_{\ell^\prime}$, 
$\ell^\pm \nu_\ell q\bar{q}^{\prime}$, 
$q_1\bar{q}^{\prime}_1 q_2\bar{q}^{\prime}_2$, plus a $\n{1}$. 
An additional possibility is the two-body mode $\n{2}\to\n{1}\h(\A)$, 
when the Higgs boson(s) is (are) light enough \cite{Gunion-Haber,Ambr-Mele-2}.
This may give rise to events with a $b\bar{b}/\tau^+\tau^-$ pair and 
missing energy and momentum ($\miss{E}$ and $\miss{p}$) in the detectors. 

The radiative $\n{2}$ decay into a photon and a lightest
neutralino \nrad\ provides a further decay channel with an even more 
interesting signature. Analytical formulas for the corresponding width
can be found in Ref.\cite{Haber-Wyler}.\footnote{In checking the results of 
Ref.\cite{Haber-Wyler}, we found that the first line in Eq.(59),
p.~281, should be more properly written as follows: 
\[G_L = 2\cos\theta_t[T_{3t}Z_i^{(-)}+e_t\tgw
Z_{i1}]+\sin\theta_t Z_{i4} \frac{m_t}{m_W\sinb}.\] In this way, it 
can also be correctly generalized to the $f=b,\tau,\ldots$ case, by
substituting $T_{3t}, e_t, \theta_t \to T_{3f}, e_f, \theta_f$.} 
Because of the higher-order coupling, this channel is characterized in 
general by rather low branching ratios (BR's). Nevertheless, a 
comprehensive study of the $\n{2}$ decay rates \cite{Ambr-Mele-2} shows 
that there are regions of the SUSY parameter space where the radiative 
decay is important and can even become the dominant $\n{2}$ decay.
 
In this paper, we  perform a detailed phenomenological analysis of the 
decay \nrad\ in the MSSM. In particular, we analyze the regions of the 
\susy\ parameter space where the radiative decay is enhanced.
Some asymptotic regimes that give rise to large radiative BR's 
have been considered in Ref.\cite{Haber-Wyler}.
In these particular cases, a considerable hierarchy is  
present among the different mass parameters of the MSSM Lagrangian,  
that by now corresponds to regions of the \susy\ parameter space either 
partly or entirely excluded by LEP searches. 
In our study, we go beyond such asymptotic scenarios and consider regions 
of the \susy\ parameter space where the parameters $M_{1,2}$ and $|\mu|$  
have values roughly included in the range $[M_Z/4, 4 M_Z]$.  
We relax the usual condition on the electroweak gaugino masses 
$M_1=\frac{5}{3}\tgwq M_2$, which holds [through one-loop renormalization 
group equations (RGE's)] when one assumes their unification at a scale 
$M_{\it GUT} \approx 10^{16}$ GeV, where the gauge couplings 
assume the same value.\footnote{Note that this not necessarily
corresponds to relaxing all the gaugino mass unification conditions. 
One can still imagine that $M_2$ and $M_3$, that is the parameters which 
correspond to non-abelian gauge groups, unify in the usual way, while 
the unification relation between $M_1$ and $M_2$ may be different 
from the usual one and unknown.} 
In this paper, we will treat $M_{1,2}$, as well as the masses of the
individual sfermions, as low-energy, free parameters. As for the 
sfermion masses, we will assume, for simplicity, some common
value (or two different values for sleptons and squarks), whenever the 
individual mass values are not particularly relevant for the analysis. 

As we will show, taking both $M_1$ and $M_2$ as free parameters can
produce new interesting scenarios beyond the ones already considered
in Ref.\cite{Ambr-Mele-2}. In that paper, we already singled out some 
regions of the parameter space where the tree-level $\n{2}$ decays are 
suppressed and BR(\nrad) is large. There, we partly misinterpreted the 
origin of the suppression in a few points of the most promising regions  
in the ($\mu, M_2$) plane, ascribing it to the particular physical 
composition of $\n{1,2}$. It will be clear from the following more complete 
analysis that in those scenarios a {\it kinematical} suppression of the 
tree-level decays can be effective, in addition to a {\it dynamical} one. 
We indeed noticed in Ref.\cite{Ambr-Mele-2}, that the two lightest 
neutralinos are nearly degenerate in the interesting cases and that their 
mass difference $(\mn{2}-\mn{1})$ grows monotonically with $\tgb$. 
Now, we will show how different mechanisms can contribute to the radiative 
BR enhancement. These scenarios will be thoroughly analyzed by a systematic 
investigation of the regions of the \susy\ parameter space where a large 
BR(\nrad) regime may be present. For instance, a {\it dynamical} suppression 
of the $\n{2}$ tree-level decays occurs when $\n{1}$ and $\n{2}$ have a 
different dominant physical composition in terms of gauginos and Higgsinos. 

 The latter scenario can have particular relevance for explaining events 
like the \eegg\ event recently observed by the Collider Detector at Fermilab 
(CDF) Collaboration at the Tevatron \cite{CDF-eegg}. That event,  
characterized by the presence of hard photons, electrons and large 
missing transverse energy, can be hardly explained within the Standard
Model (SM). A possible solution to this puzzle can be found within
the minimal \susy\ models, by interpreting the CDF event as a result 
of selectron- (or chargino-) pair production, provided a large BR for 
the neutralino decay into a LSP and a photon is predicted.  
More generally, the presence of a large radiative neutralino decay BR is 
crucial to obtain, within \susy\, high rates for final states associated to 
a signature of the kind $\gamma\gamma + X + \miss{E}$, where 
$X = \ell^+\ell^{(\prime)-}, \; q\bar{q},\; \ldots$ or {\it nothing}. 

Two different \susy\ ``models'' have been proposed up to now to explain
the \eegg\ CDF event \cite{UM-Paper,SLAC-Paper,UM-Paper2}.\footnote{As 
we were completing this paper, three other papers appeared which discuss 
the CDF event in various contexts \cite{Other_eegg}.}
The first one arises within theories with low-energy supersymmetry
breaking, where the breaking is transmitted to the visible sector by 
nongravitational interactions \cite{GMSB}. In such a scenario, the 
gravitino $\tilde{G}$ turns out to be naturally the LSP and, if 
light enough (i.e., for $m_{\tilde{G}} \ltap 1$ keV), it can couple to  
standard \susy\ particles strongly enough to be of relevance for collider 
phenomenology \cite{Fayet}. If one assumes that the lightest 
{\it standard} \susy\ particle is still the lightest neutralino and 
$R$-parity is conserved, all the heavier \susy\ particles produce, at
the end of their decay chain, (at least) one $\n{1}$. 
Then, $\n{1}$  decays radiatively, $\n{1}\to \tilde{G}\gamma$, through its 
photino component, if the latter is not tuned to zero \cite{UM-Gravitino}. 
Assuming the $\n{1}$ radiative decay occurs well into the detector 
(i.e., close enough to the main vertex, which requires $m_{\tilde{G}} \ltap 
250$ eV), such a model  can provide a satisfactory explanation of the \eegg\ 
event \cite{UM-Paper,SLAC-Paper}. 
Once low-scale \susy\ breaking and very light gravitino
scenarios are trusted, the presence of a large-BR $\n{1}$ radiative decay 
and the consequent signature of hard and central photons and missing energy 
are almost automatic \cite{UM-Paper,SLAC-Paper,UM-Gravitino}. 
Although quite general, such a hypothesis does not allow to predict 
much about the \susy\ parameters apart from $m_{\tilde{G}}$. 
In particular, an interpretation of the CDF event within this framework 
can single out some ranges for the physical masses of the involved 
particles only on the basis of a careful analysis of the kinematical 
characteristics of the event \cite{UM-Paper,UM-Paper2}. 
No specific statements about the values of the parameters in the \susy\
Lagrangian ($M_1$, $M_2$, $\mu$, $\tgb$, etc.) and, hence, no detailed 
predictions of the general related collider phenomenology can be achieved. 

Somehow opposite is the situation if the CDF event (or a general 
$\gamma\gamma+X+\miss{E}$ event) is explained within the MSSM,
where the gravitino is heavy, the lightest neutralino is the LSP and 
the hard photons and the missing energy are due to the one-loop  
$\n{2}\to\n{1}\gamma$ decay \cite{UM-Paper,UM-Paper2}. 
In this case, a certain adjustment of the MSSM parameters is required 
(both in the {\it dynamical}- and in the {\it kinematical}-enhancement
scenarios) in order to get large radiative BR's and large rates for events
with hard central photons and missing energy. 
Also, if the hard photons are emitted by rather soft $\n{2}$'s,
the {\it dynamical} enhancement is the most effective mechanism 
in this respect. Then, one is in general able to select rather 
narrow ranges of the \susy\ parameters, if the $\gamma\gamma+X+\miss{E}$ 
events are interpreted in this framework \cite{UM-Paper,UM-Paper2}. 
Hence, such a framework can be quite predictive also about the 
\susy\ collider phenomenology that should show up in the future. 

In this work, we intend to investigate the latter hypothesis in a  
general framework. We simply look for regions in the usual MSSM parameter 
space where the \nrad\ decay has sizeable BR's. 
We find that in order to have a large BR(\nrad) (up to about $100\%$)
one needs in general $\tgb \simeq 1$ and/or $M_1 \simeq M_2$, in
addition to $\mu<0$.\footnote{In this paper, we follow the same convention 
as in Ref.\cite{Haber-Kane} for the sign of $\mu$. We also assume 
$M_{1,2} \ge 0$ and $\tgb \ge 1$. Note that large values of $\tgb$ 
($\gtap 60$) are disfavored by a radiative electroweak symmetry breaking in 
the MSSM \cite{RSB}.} This is a quite general requirement, while further 
conditions on the gaugino mass parameters $M_1$, $M_2$, the Higgsino mass 
$|\mu|$, and $\tgb$ can guarantee either a {\it dynamical} or a 
{\it kinematical} enhancement of BR(\nrad). These two possibilities will 
give rise to rather different spectra for the emitted photons. 

The effects on BR(\nrad) of varying all the masses in the sfermion sector 
is also considered. In particular, the characteristics of the top-squark 
sector are quite relevant for the radiative decay width
\cite{Haber-Wyler,Old-Radiative,Bartl-Neu}. 
In our previous studies \cite{Ambr-Mele-1,Ambr-Mele-2}, we assumed 
all the left and right squarks degenerate in mass, in order to simplify  
the multiparameter dependence of the neutralino phenomenology. 
We also neglected the effects of a possible top-squark mixing. 
Here, we will examine the general case and we will see that the behavior
of the BR(\nrad) can be affected by the top-squark sector parameters in
different ways when the radiative BR is enhanced {\it dynamically} or 
{\it kinematically}.

The outline of the paper is the following.
In Sec.~2, we review the theory of the radiative neutralino decay in the 
MSSM and fix the notations. We also introduce the possible scenarios where 
a large BR(\nrad) regime can arise. In Secs.~3 and 4, respectively,
we go through the \susy\ parameter regions where a {\it dynamical} and a 
{\it kinematical} enhancement of the radiative decay can take place. 
In Sec.~5, we perform a  numerical analysis of the radiative BR in the 
relevant parameter regions. Finally, in Sec.~6, we study the top-squark 
sector influence on the radiative neutralino decay. 
In Sec.~7, we draw our conclusions. 

\section{Enhanced BR(\nrad) regimes}

\begin{figure}[h]
\centerline{
\epsfxsize=\textwidth 
\epsffile{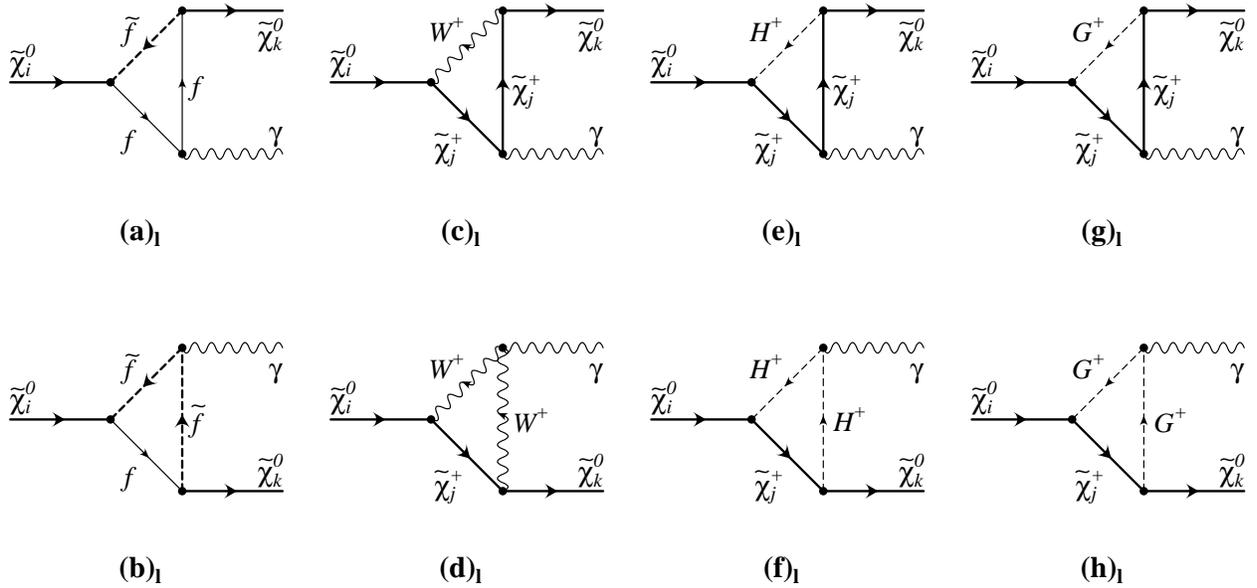} 
}
\caption{Feynman diagrams for the radiative neutralino decay 
$\protect\n{2}\to\protect\n{1}\gamma$, in the gauge of
Ref.\protect\cite{Haber-Wyler}. For each graph shown, there is a further
one with clockwise circulating particles in the loop.}
\label{feyrad}
\end{figure}

The radiative neutralino decay receives contributions, in a convenient
gauge \cite{Haber-Wyler}, from 16 graphs with all the charged (both
\susy\ and non-\susy) standard particles flowing in the loop. 
The corresponding Feynman diagrams are displayed in Fig.~\ref{feyrad}. 
As a result, \nrad\ is a very interesting process in itself, since it is
influenced, to some extent, by all the parameters and sectors of the
MSSM. However, being of higher order with respect to the main tree-level 
decays, the channel \nrad\ typically has a BR not larger than a few percent.
In previous studies \cite{Old-Radiative}, the possibility 
that BR(\nrad) gets large in the special case $\n{2}\simeq\hino{}$, 
$\n{1}\simeq\phino$ has been considered, according to approximate 
formulas for the matrix element of the process, including only the main
contributions from the $\Wpm/\cpm{}$ and $t/\tilde{t}_1$ loops. 
In Ref.\cite{Haber-Wyler}, after the full calculation of the matrix
element and the decay width, two examples of scenarios with 
large BR(\nrad) are given, in two different limits. {\it Asymptotic} 
values of the relevant \susy\ parameters, according to which the actual 
radiative processes are 
$\phino\to\hino{}\gamma$ and $\hino{1}\to\hino{2}\gamma$,
are considered (see below). 
In Ref.\cite{Ambr-Mele-2}, by using the full calculation and assuming 
the gaugino-mass unification, we stress the presence of {\it non-asymptotic} 
regions of the parameter space, where the radiative process is still enhanced. 
In the following more general approach, we  show that all the above 
scenarios are just particular realizations of the two main enhancement 
mechanisms for the radiative decay BR.

The naive expectation that the BR for \nrad\ is negligible with respect to
the BR's for the tree-level neutralino decays is not realized whenever the
latter channels are suppressed for reasons that either do not affect or
affect to a minor extent the radiative process.
This can happen basically in two cases:
\begin{description} 
\item[a) Dynamical suppression.] ~\\    
Neutralinos are in general superpositions of gauginos ($\phino$ and
$\Zino$) and Higgsinos ($\hino{a}$ and $\hino{b}$).\footnote{For the
neutralino/chargino sectors, we use notations similar to 
Refs.\cite{Bartl-Neu,Bartl-Matr}. In particular, 
for the neutralino mixing matrix, we use the basis ($\phino$,
$\Zino$, $\hino{a}$, $\hino{b}$), instead of ($\tilde{B}$, $\tilde{W}_3$,
$\hino{1}$, $\hino{2}$), used  in Ref.\cite{Haber-Kane}.
This choice is particularly suitable for our purposes.}  
The couplings of sfermions to neutralinos involve only
the gaugino components [apart from terms $\approx (m_f/M_W)$,
where $m_f$ is the mass of the standard fermion also entering the vertex], 
while the $\Z$ only couples through Higgsinos \cite{Haber-Kane,Bartl-Neu}.
Then, the direct tree-level decays $\n{2}\to\n{1}f\bar{f}$ require either
simultaneous gaugino components in both $\n{1}$ and $\n{2}$
(for the sfermion-exchange process) or simultaneous Higgsino components 
(for the $\Z$-exchange process).
This is partly true even when the exchanged particles are on their
mass shell, that is when the two-body channels $\n{2}\to f\tilde{f}$ 
and/or $\n{2}\to\Z\n{1}$ are open.   
The above requirement does not hold for the radiative decay, since in 
general both the gaugino and Higgsino components of neutralinos are
involved in each graph of Fig.~\ref{feyrad}, 
apart from the diagrams (a)-(b) for a massless fermion $f$. 
Hence, whenever $\n{1}$$(\n{2})$ is mainly a gaugino while $\n{2}$$(\n{1})$ 
is dominated by the Higgsino components, the tree-level $\n{2}$ width
for direct decays falls down and the BR(\nrad) is enhanced.
For pure gaugino/Higgsino states, BR(\nrad) can reach 100\%.
In particular regions of the parameter space, this  picture 
can be modified by the presence of a light chargino. 
Indeed, the cascade decays  
$\n{2}\to\cpm{1}(\to\n{1}f_1\bar{f}^{\prime}_1)f_2\bar{f}^{\prime}_2$,
when kinematically allowed, can take place even for different physical 
composition of the two neutralinos, through $\Wpm$-exchange graphs 
involving both $\tilde{W}_3$'s and $\hino{}$'s.  
Note that a different dynamical suppression (that we call {\it reduced} 
dynamical suppression) of the tree-level direct decays can take place when 
the sfermions are heavy and the $\Z$-exchange channel is dynamically 
suppressed by the presence of a dominant gaugino component in at least one 
of the two neutralinos. The latter case will be of some relevance in our 
following numerical analysis. 

As for the decays into Higgs bosons, when 
$m_{\h} (m_{\A}) < (\mn{2} - \mn{1})$,
the two-body channels $\n{2}\to\n{1}\h(\A)$, open up too.
Naively, the latter do not seem to suffer from any dynamical 
suppression, since the MSSM predicts the vertex $\Zino\hino{i}\h(\A)$. 
However, we will see that an effective dynamical suppression can 
be achieved when one of the two lightest neutralinos is dominated by a 
$\phino$ component (not just any gaugino!). In this case, due to the
absence of the $\phino\hino{i}\h(\A)$ vertex, the neutralino decays into 
Higgs bosons are depleted as well, and the BR(\nrad) can still be
non-negligible. 
\item[b) Kinematical suppression.] ~\\    
When $\n{2}$ and $\n{1}$ tend to be degenerate in mass, the widths for the 
different $\n{2}$ decay channels approach zero differently as the quantity  
$\protect\epsilon \equiv (1 - \mn{1}/\mn{2})$ vanishes. 
For the radiative decay, one has \cite{Haber-Wyler} 
\beq 
\Gamma(\n{2}\to\n{1}\gamma) = 
 \frac{g^2_{\n{1}\n{2}\gamma}}{8\pi} 
 \frac{(\mn{2}^2-\mn{1}^2)^3}{\mn{2}^5}
\begin{array}{c}\phantom{.} \\
\widetilde{\ \sss \epsilon \to 0 \ } \end{array} 
 \frac{g^2_{\n{1}\n{2}\gamma}}{\pi}\mn{2}\epsilon^3 \; ,  
\label{Gamma_Rad}
\eeq 
where $g_{\n{1}\n{2}\gamma} \propto eg^2/16\pi^2$ is an effective coupling
arising from the one-loop diagrams in Fig.~\ref{feyrad} 
(in general a complicated function of all the masses and couplings 
to neutralinos of the particles circulating in the loops).  

On the other hand, the three-body direct tree-level decays receive
contributions from either $\Z$-exchange graphs or sfermion-exchange 
graphs. The former ones, involving the Higgsino components only, in the 
limit of small $\epsilon$ and massless standard fermions $f$, lead 
to a total width of \cite{Haber-Wyler,Bartl-Neu}
\beq 
\sum_f\Gamma(\n{2}\to\n{1}f\bar{f})_{\Z-{\rm exch.}} 
\begin{array}{c}\phantom{.} \\
\widetilde{\ \sss \epsilon \to 0 \ } \end{array} 
\frac{g^4 C_w C_{\hino{}}}{\pi^3} \frac{\mn{2}^5}{M_Z^4} \epsilon^5 \; , 
\label{Gamma_Z} 
\eeq 
where $C_w \approx 10^{-2}$ and $C_{\hino{}}$ is a number
$\le 1$, depending on the Higgsino content of the neutralinos
(for pure Higgsinos: $f_{\hino{}} = 1$). 
Equation (\ref{Gamma_Z}) implies a sum over colors and five (six) 
flavors of final-state quarks (leptons). 
A similar behavior is found for the sfermion-exchange graphs.\footnote{
As can be inferred, for instance, from the treatment of analogous 
channels for the gluino decay to a photino plus a quark pair, in 
Ref.\protect\cite{Haber-Kane}.}
One has, for a single channel into a given $f\bar{f}$ pair, mediated 
by a left- or right-sfermion  
\beq 
\Gamma(\n{2}\to\n{1}f\bar{f})_{\tilde{f}-{\rm exch.}}  
\begin{array}{c}\phantom{.} \\
\widetilde{\ \sss \epsilon \to 0 \ } \end{array} 
\frac{g^4 C^{L,R}_{\tilde{A}}}{\pi^3} 
\frac{\mn{2}^5}{m_{\tilde{f}_{L,R}}^4}  \epsilon^5 \; , 
\label{Gamma_sf} 
\eeq 
where $C^{L,R}_{\tilde{A}}$ is typically $\approx 10^{-2} - 10^{-1}$, 
but can be slightly larger or much smaller, depending on the 
gaugino content of the neutralinos and on the specific channel considered. 
Yukawa couplings of the Higgsino components to $f\tilde{f}$, 
as well as L-R mixings for the exchanged sfermions $\tilde{f}$, 
are here coherently neglected, since we work in the massless 
$f$ approximation and top (s)quarks are not involved in the 
problem, $m_t$ being too heavy.   
As for the interference term of the $\Z$- and sfermion-exchange 
graphs, we expect of course the same fifth-power behavior as in 
Eqs.(\ref{Gamma_Z}) and (\ref{Gamma_sf}).  
This implies that the ratio of the direct tree-level and the radiative 
decay widths tends to vanish as $\epsilon^2$, when $\epsilon\to 0$ 
or $\mn{2} \to \mn{1}$.    
Note also that the presence of the small number $C_w$ in Eq.(\ref{Gamma_Z}) 
can partly compensate the additional factor of order $\alpha_{\rm em}$ in 
Eq.(\ref{Gamma_Rad}). Hence, for neutralino masses of interest for 
present/near-future collider physics and for this work 
(where $\mn{1,2} \approx M_Z$), when ($\mn{2} - \mn{1}) \sim 10$ GeV
it is already possible for the radiative decay BR to receive a
substantial factor of enhancement $\approx 10^2$, especially if the 
sfermion-exchange channels are suppressed for some dynamical reason. 

At this point, it is important to stress that, when neutralinos are 
degenerate within less than about 10 GeV, the asymptotic formulas given 
above for \mbox{$\Gamma(\n2\to\n{1}f\bar{f})_{\Z-, \ \tilde{f}-{\rm exch.}}$} 
are no longer a valid approximation when $m_f$ cannot be neglected with 
respect to ($\mn{2} - \mn{1}$), as in the case, e.g., 
$f = b$ or $\tau$.\footnote{One should, instead, use the full formulas, 
including finite $m_f$ effects, to get the correct trend of the width as
a function of the mass difference 
$\Delta m = \mn{2} - \mn{1} - 2 m_f \equiv \mn{2} (1 - \epsilon_f)$ 
between the initial and final states, as it approaches (or gets even
smaller than) $m_f$. One expects that the simple fifth-power behavior 
in terms of $\epsilon \simeq \epsilon_f$ in Eqs.(\ref{Gamma_Z}) and
(\ref{Gamma_sf}) will be gradually spoiled while entering the
``intermediate'' kinematical region where $\epsilon_f$ becomes 
sensitively different from $\epsilon$. Here, one would have to
consider much more involved formulas for the widths as functions of  
$\epsilon_f$. Eventually, when $\Delta m$ becomes much smaller 
than $m_f$ too, a simple asymptotic regime is again effective, 
but in terms of $\epsilon_f$, and possibly different from the former
fifth-power-like one.} 

However, in the subtle kinematical regions around the various 
heavier $m_f$ ``thresholds'', the $\n{2}\to\n{1}f\bar{f}$ channels with 
lighter $f$'s will dominate over the ones with $f=b, \tau, \ldots$. 
Hence, neglecting $m_f$ everywhere and using Eq.(\ref{Gamma_Z}) and 
Eq.(\ref{Gamma_sf}) summed over all non-top flavors, will have the net 
result of an overestimate of the total width for the $\n{2}$ tree-level
decays and an underestimate of BR(\nrad). The latter simplified
treatment of the problem (which is the one we will adopt in all the 
following numerical analyses), however, is here justified by at least  
two good motivations. First, we have already seen that for $\mn{1,2} 
\approx M_Z$ and when $\mn{2} - \mn{1} \sim 10$ GeV $\simeq 2 m_b$, 
that is close to the heaviest $m_f$ ``threshold'', the radiative 
decay BR can already receive a $\approx 10^2$ enhancement factor. This
factor can be extracted directly from Eqs.(\ref{Gamma_Z}), 
(\ref{Gamma_sf}), which in this region are still a reliable approximation.
Second, considering values for neutralino masses and the difference 
($\mn{2} - \mn{1}$) with a precision at the level of a few GeV or less, 
for a given set of input parameters, is not fully sensible when
neglecting radiative-correction effects on the spectrum, as we do in 
this paper. (A more extensive discussion of this problem can be found at 
the end of this section.)  

 Having the above caveats in mind, we will not be concerned in the 
rest of the paper with fine behavior and subtle kinematical effects for 
neutralino mass differences at the level of several GeV or less. 
Nevertheless, we expect that the numerical analyses to be performed in the 
following are valid in most of the interesting cases. Even for 
very small ($\mn{2} - \mn{1}$), our identification 
of the parameter space regions where this can happen, and the result of a
large BR(\nrad) (with approximately the numerical value we will indicate) 
keep holding. 

Regarding the cascade decays through light charginos, they are at least as 
kinematically suppressed as the normal direct three-body decays.
Indeed, when $\epsilon \to 0$, the width of each of the two steps of the 
cascade decay has a fifth-power asymptotic behavior for 
$\epsilon_{1,2} \to 0$, where $\epsilon_{1,2} = (1 - \mc{1}/\mn{2})$, 
$(1 - \mn{1}/\mc{1})$. Furthermore, $\epsilon_{1,2} < \epsilon$, since 
$\mn{1} < \mc{1} < \mn{2}$ must hold, for the cascade to take place. 
On the other hand, some of the channels for this class of decays
will not be suppressed by possibly small couplings like the $C$'s, in 
Eqs.(\ref{Gamma_Z}) and (\ref{Gamma_sf}). 
As for situations with very small neutralino/chargino mass differences, 
similar remarks as for the direct case above hold here. 
 
Finally, the two-body decay into Higgs bosons cannot take place when 
the mass difference between the two lightest neutralinos is less than a few 
tens of GeV, because of the current experimental limits on $m_{\h}$ and 
$m_{\A}$ \cite{PDG}. 
\end{description} 

The conditions {\bf a)} and {\bf b)} can be translated into requirements 
on the \susy\ parameters $\tgb$, $\mu$, $M_1$ and $M_2$, which set the mass 
matrix of the neutralino sector. \\ 
The tree-level neutralino mass matrix reads, in the convenient basis \\ 
$[-i\phino, \; -i\Zino, \; \hino{a}=\hino{1}\cosb - \hino{2}\sinb, 
\; \hino{b}=\hino{1}\sinb + \hino{2}\cosb]$
\beq 
{\cal M}_{\n{}} = \left[ \begin{array}{cccc} 
  M_1 \cthwq + M_2 \sthwq & (M_2 - M_1) \sthw\cthw      & 0    & 0        \\ 
 (M_2 - M_1)\sthw\cthw    & M_1 \sthwq + M_2 \cthwq     & M_Z  & 0        \\ 
    0      &      M_Z     &         \mu \sindb          &  -\mu \cosdb    \\ 
    0      &      0       &        -\mu \cosdb          &  -\mu \sindb    \\ 
\end{array} \right] \; . 
\label{Neu-Matrix} 
\eeq 
It is easy to recognize in Eq.(\ref{Neu-Matrix}) two 
$2\times 2$ blocks, which correspond to: \\ 
(i) the gaugino mass terms, parameterized by $M_1$ and $M_2$ and mixed 
by the weak angle. Their source is soft \susy\ breaking. \\ 
(ii) the Higgsino mass terms, parameterized by $\mu$ and $\tgb =
v_2/v_1$, whose source is a \susy\ term in the MSSM Lagrangian, 
which mixes the Higgs doublets. \\
Then, there are only two off-diagonal entries non included in the two 
$2 \times 2$ blocks, corresponding to $\hino{a}$-$\Zino$ mixing
terms and equal to $M_Z$, which come from the $H\hino{}\Zino$ couplings
and the \susy\ Higgs mechanism. 
As a consequence, apart from the asymptotic cases where $M_{1,2}$ and/or
$|\mu|$ are much larger than the $\Z$ mass, it is not possible to have 
either a pure $\Zino$ or a pure $\hino{a}$. Hence, whenever a neutralino
has a sizeable $\Zino$ component, it must have a sizeable
$\hino{a}$ component as well (and vice versa). 
This means that a neutralino can be a pure gaugino only when it is a 
photino, and a pure Higgsino can only be of the $\hino{b}$ type 
(sometimes called ``symmetric Higgsino'', with notation $\hino{S}$). 
Note also that when $M_1 = M_2$ (or $\tgb = 1$) the off-diagonal terms
within the $2\times 2$ gaugino (or Higgsino) block disappear. The 
limits $(M_1 - M_2) \to 0$ and $\tgb \to 1$ will be crucial for 
the enhancement of the neutralino radiative decay. 

The outcome of the neutralino mass matrix (\ref{Neu-Matrix}) 
in terms of the neutralino physical compositions and mass eigenstates 
has been extensively studied in Ref.\cite{Bartl-Matr}.
We use the results of that analysis and concentrate here
on what is relevant in order to realize either a dynamical or a
kinematical enhancement of the \nrad\ decay.

As already mentioned, some {\it asymptotic} regimes, where \nrad\ is 
enhanced, were anticipated in Ref.\cite{Haber-Wyler}. 
Some enhancement is expected in the following two cases:
\begin{description} 
\item[(i) Light-Neutralino Radiative Decay] 
($M_1, M_2, |\mu| \ll M_Z$). \\  
Then, a dynamical \nrad\ enhancement is realized, since $\n{1} \to \phino$ 
and $\n{2} \to \hino{b}$ or vice versa. In Ref.\cite{Haber-Wyler},  
the limit $\mu = 0$, $\phino\to\hino{}\gamma$ is treated analytically.
Such small values of the parameters are not yet excluded by LEP data,  
provided $1 < \tgb \ltap 2$ (what is called sometimes 
``light Higgsino-gaugino window'' \cite{Feng}). 
In fact, in this particular region, a number of things happens:\\ 
(a) when $|\mu|/M_Z, M_2/M_Z \to 0$, the chargino mass 
generally satisfies the current LEP lower bounds; \\ 
(b) if also $M_1/M_Z \to 0$, the neutralino mass eigenstates are: 
$\phino$, $\hino{b}$ (with mass eigenvalue $\to 0$) and the symmetric
and antisymmetric combinations of $\Zino$ and $\hino{a}$ (with mass 
$\to M_Z$). Then, when $\tgb \to 1$, the light $\hino{b}$
decouples and the neutralinos can only interact with the $\Z$ boson
through the vertex $\Z\hino{a}\hino{b}$, which is largely suppressed by 
the phase space at LEP1 energies, since $(m_{\hino{a}}+m_{\hino{b}})
\simeq M_Z$. \protect\\ 
As a result, the data on the $\Z$ peak can hardly constrain this
particular region. 
For instance, in Ref.\cite{L3}, the analysis is performed by considering
the bounds on $\Gamma(\Z\to\n{1}\n{1})$ from the invisible $\Z$ width 
and those on $\Gamma(\Z\to\n{1}\n{2}, \; \n{2}\n{2})$ from the direct search
of neutralinos. However, only the decays 
$\n{2}\to\n{1}Z^{(*)}\to\n{1}f\bar{f}$ are fully taken into account,
while the radiative decay, although generally dominant in this region, is 
not properly stressed. In the analysis of Ref.\cite{Feng}, a tighter BR 
bound from the radiative neutralino decay has been included as well as the
effects of data taken at $\sqrt{s}$ above $M_Z$. 
In spite of that, part of the region in the ($\mu, M_2$) plane with 
$M_2, \; |\mu| \ltap 10$ GeV still survives for $\mu < 0$ and $\tgb$ 
close to 1.   

This region is wider when $M_1$ is taken as a free parameter and 
allowed to be quite larger than $M_2$. 
Some significant improvements in probing the above region  
could come from a careful analysis of the data of the recent LEP short 
runs at $\sqrt{s} = 130$-136, 161 and 172 GeV and the future runs around  
$\sqrt{s} = 190$ GeV. If the light gaugino-Higgsino scenario is 
realized, all charginos and neutralinos are expected to have masses in 
the kinematical reach of LEP2 and should not escape detection. 
\item[(ii) Higgsino-to-Higgsino Radiative Decay] 
($|\mu|, M_Z \ll M_1, M_2 \approx$ TeV). \\  
This corresponds to a particular asymptotic case of the kinematical 
\nrad\ enhancement. Indeed, in this situation the two lightest neutralinos 
have nearly degenerate masses close to $|\mu|$, and are both almost pure 
Higgsinos. Hence, the direct tree-level decays can only
proceed through $\Z$-exchange graphs and the ratio between the
corresponding width and the $\hino{1,2} \to \hino{2,1} \gamma$ width 
can be obtained from Eqs.(\ref{Gamma_Z}) and (\ref{Gamma_Rad}), 
and is independent of the sfermion masses.  
In a sense, this is an 
optimization of the {\it kinematical}-suppression mechanism, since 
such a dynamical suppression of the contributions from the 
sfermion-exchange diagrams allows the fifth-power asymptotic 
behavior of the tree-level decay width and the $\epsilon^2$ enhancement 
of BR(\nrad) to get always effective for values of ($\mn{2} - \mn{1}$) 
at the level of 10 GeV or less [cf. \mbox{Eqs.(1--3)]}. 
On the other hand, in the present case, the factor $C_{\hino{}}$ in the 
numerator of Eq.(\ref{Gamma_Z}) is close to its maximum 1, hence depleting 
a bit the radiative-BR enhancement [cf. Eq.(\ref{Gamma_sf})].    
In the asymptotic limit \cite{Haber-Wyler}, one finds 
\beq 
\frac{\Gamma(\hino{1}\to\hino{2}\gamma)}{\Gamma(\hino{1}\to\hino{2}f\bar{f})}
\sim 0.3 C^2 \alpha_{\rm em} 
\left[ \frac{M_1 M_2}{\mu (M_1 + M_2 \tgwq)} \right]^2 \; ,  
\label{Hino_case}
\eeq 
where $C$ is a number of order unity weakly dependent on the ratio 
$M_W^2/\mu^2$. Here, for large $M_{1,2}$, values of $|\mu| \ltap M_Z/2$ 
are generally already excluded by LEP1-LEP1.5 data, since they lead to 
chargino masses lighter than 45--50 GeV.
Hence, Eq.(\ref{Hino_case}) tells us that one needs {\it very} large 
values of $M_{1,2}$ to get a significant enhancement of the radiative 
decay through this mechanism. In addition, in this region, $\n{2}$ can 
often decay through cascade channels into a lighter chargino with a
non-negligible branching fraction. Numerically, by using the full 
formulas, we checked that BR(\nrad) 
$\ltap 10\%$ always for $M_{1,2} \ltap 2-3$ TeV, and 
$|\mu| \gtap 45$--50 GeV.  
Furthermore, it turns out that to have a sizeable radiative decay in
this case (that is to suppress both direct and cascade tree-level
decays), one always has to enforce the condition $\mn{2} - \mn{1} 
\ltap 2-3$ GeV, which drastically restricts the photon energy and 
also corresponds to a critical kinematical region, as discussed above. 

Also note that, for nearly degenerate Higgsinos, the radiative
corrections may actually spoil the enhancement mechanism or, at least,
render the tree-level analysis rather inaccurate, even if one 
takes into account finite $m_f$ effects. For instance, the Higgsino-mass 
splittings in this region can receive radiative corrections as large as 
their tree-level values [i.e., up to $\pm$(5--10) GeV], if the mixing 
between the two top-squarks is large~\cite{Giudice-Pomarol}. 
\end{description} 

Now, we want to extend the quoted studies by analyzing the more general 
framework where either a dynamical or a kinematical \nrad\ enhancement 
can be realized, without assuming any particular hierarchy between the 
\susy\ parameters and relaxing the usual unification condition on the 
gaugino masses.
We will not concentrate on particular limits of the \susy\ parameters 
such as $M_1$ and/or $M_2 \to 0$ and/or $|\mu| \to 0$, since they have 
been either already excluded by LEP data or discussed above. 

In the following, we will neglect the effects of the radiative
corrections on the neutralino mass matrix elements. In 
Ref.\cite{Pierce-Papa}, the full calculation has been carried out.
Regarding the radiative corrections to the neutralino mass 
eigenvalues, they are found to be generally at the level of 3-8\% and
of the same sign (positive) for all the mass eigenvalues. 
Only occasionally, the lightest neutralino mass can receive larger
corrections. No conclusion can be easily extracted from that analysis 
about whether or not and how much the radiative corrections may change the
composition of a neutralino eigenstate and, in particular, to what
extent, for instance, an almost pure photino at the tree level, could 
turn out to be a more mixed state at the one-loop level. 
Our scenarios for an enhanced radiative-decay regime rely on a certain 
amount of adjustment between different \susy\ parameters in order to get 
either pure compositions for the eigenvectors or degeneracy for the 
eigenvalues of the neutralino mass matrix. 
As for the kinematical-enhancement mechanism, we will show that 
a mass difference $\mn{2} - \mn{1} \sim 10$ GeV, 
with $\mn{1,2} \approx M_Z$, is in general sufficiently small
to get a BR(\nrad) of order 40\% or more. 
Since the higher-order corrections to the neutralino 
masses have generally a fixed sign, one can expect a 
common shift of the different masses, 
while the relative mass differences change only slightly. 
Thus, we expect that the kinematical suppression 
keeps almost unchanged  for $\mn{2} - \mn{1} \sim 10$ GeV and
that our treatment substantially holds even after
radiative corrections, with a possible slight redefinition of the 
interesting regions in the \susy\ parameter space. 
Concerning the dynamical-enhancement mechanism, similar arguments may
be used, although in this case there are less clues 
to guess the effects of radiative corrections. 
However, we will see that the amount of
parameter adjustment required for the mechanism to be effective  
is not very large. For instance, there are significant
regions in the parameter space where 
$|\langle\n{1,2}|\hino{b}\rangle|^2$  and/or 
$|\langle\n{2,1}|\phino\rangle|^2$ 
are {\it only} about 0.8 and the dynamical-suppression
is still effective, with the BR(\nrad) of order 50\% or more.
Hence, an adjustment of the parameters at the
level of 20\% should survive the inclusion of the radiative corrections.

\vspace{0.3cm} 

An additional remark is due for the case of the kinematical enhancement 
mechanism. Since the latter arises from situations where the two lightest 
neutralinos are close in mass, a too strong degeneracy may prevent the 
experimental detection of the $\n{2}$ decay, due to the emission of too 
soft photons. In general, one can ensure the presence of a useful experimental 
signature and the phenomenological relevance of the neutralino radiative decay 
by requiring $\mn{2} - \mn{1} \gtap 10$ GeV. This, of course, effectively  
depletes the actual BR(\nrad) that can be achieved by the kinematical 
mechanism in a real experimental framework. 
On the other hand, if the available c.m. energy is large enough, the 
$\n{2}$ can receive a sizeable boost, even when $(\mn{2} - \mn{1})$ is 
very small (for instance, when produced in association with a $\n{1}$ at 
LEP2, or with a $\cpm{1}$ at the TeVatron).  In order to assess to what 
extent this is true, one can take into account that, assuming an isotropic 
radiative decay of the produced $\n{2}$ (that is neglecting spin-correlation 
effects), the resulting photon has a flat energy distribution in the 
laboratory, with end points 
\[  E(\gamma)_{\rm min,max} = 
\left( \frac{E_2}{\mn{2}} \mp \sqrt{\frac{E_2^2}{\mn{2}^2} - 1} \right)
\left( \frac{\mn{2}^2 - \mn{1}^2}{2\mn{2}} \right) \; , \] 
where $E_2$ is the production energy of the $\n{2}$ in the laboratory. 

\section{Dynamical \protect\nrad\ enhancement} 
\noindent
As already seen, requiring one pure gaugino and one pure Higgsino 
eigenstate from the matrix (\ref{Neu-Matrix}) implies both
\beq 
M_1 = M_2 
\label{cond1}
\eeq
and 
\beq 
\tgb = 1 \; . 
\label{cond2}
\eeq 
One then has a pure $\phino$ with mass $M_1 (=M_2)$ 
and a pure $\hino{b}$ of mass $-\mu$ in the neutralino spectrum.
In this limit, the other two neutralinos are mixtures of 
$\hino{a}$ and $\Zino$ with mass eigenvalues (including their 
sign)\footnote{Of course, the physical neutralino 
masses are always positive, but the sign of the mass eigenvalue has 
its own physical meaning, being connected to the neutralino $CP$ quantum 
numbers and entering the expressions, in the basis we adopt, of the Feynman 
amplitudes for processes involving neutralinos 
\cite{Haber-Kane,Bartl-Neu,Bartl-Matr,Neu-CP}.} \cite{Bartl-Matr}
\beq 
m_{\hino{a}-\Zino}^{(\pm)} = 
\frac{1}{2} \left[M_2 + \mu \pm \sqrt{(M_2 - \mu)^2 + 4 M_Z^2} \right] 
\label{mhz} 
\eeq 
and mixing angle
\beq 
\left\{ \begin{array}{c} \sin\phi \\ \cos\phi \end{array} \right\} 
= \frac{1}{\sqrt{2}} 
\left[ 1 \pm \frac{M_2 - \mu}{\sqrt{(M_2-\mu)^2 + 4 M_Z^2}} \right]^{1/2} \; .
\label{hzmix} 
\eeq 
Requiring that the pure states correspond to the lightest neutralinos, 
$\n{1}$ and $\n{2}$, the absolute values of both the eigenvalues in 
Eq.(\ref{mhz}) have to be larger than both $M_1$ (or $M_2$) and $|\mu|$.
In the parameter space not yet excluded by the LEP data, this can be 
achieved only if 
\beq 
\mu<0 \; .
\label{cond3} 
\eeq 
Indeed, when $\mu$ is positive, the smallest absolute value 
in Eq.(\ref{mhz}) corresponds to choosing the negative sign
before the square root. It is then sufficient to look at $m_{H-Z}^{(-)}$. 
The latter is always smaller (greater) than both $\mu$ and $M_2$, 
whenever $\mu$ and $M_2$ are larger (smaller) than $M_Z/2$. 
On the other hand, if $\mu < M_Z/2 < M_2$ or $M_2 < M_Z/2 < \mu$, 
$m_{H-Z}^{(-)}$ can still (but not necessarily has to) be large enough 
to allow the mass ordering needed for the dynamical suppression.
However, values of $M_2$ and/or $|\mu| \ltap M_Z/2$, with $\mu$ positive, 
are generally excluded by LEP data. This is true either because of the 
chargino-mass bound, or because of the direct searches of neutralinos,
even without gaugino-mass unification assumptions, and even in that
window with very small $|\mu|$ and $M_2$ we treated above, for any $\tgb$.  
As a result, the dynamical enhancement can be present only for $\mu < 0$.
Note that the eigenvalue $m_{H-Z}^{(-)}$ corresponds to a massless
neutralino when $\mu = M_2 = M_Z$. 

Quite different is the situation for $\mu < 0$. Points in the \susy\ 
parameter space where $m_{H-Z}^{(\pm)}$ are both heavier than $|\mu|$
and $M_2$ do exist for small values of $|\mu|$ and/or $M_2$.
This is also true for large values of $|\mu|$ and $M_2$, i.e. far away 
from the LEP exclusion region. For instance, let's examine the case 
$\mu = -M_Z$. In this case, whenever $0 \le M_2 < M_Z/2$, $|m_{H-Z}^{(+)}|$ 
is always less than $|\mu|$ and corresponds to the next-to-lightest 
neutralino, the lightest one being the photino with mass $M_1 = M_2$. 
Then, no dynamical enhancement can take place. For 
$M_Z/2 < M_2 < M_Z(1+\sqrt{3})/2$, the neutralino eigenstates corresponding 
to $m_{H-Z}^{(\pm)}$ are always the two heaviest ones. Then, the radiative 
decay $\hino{b} \to \phino\gamma$, for $M_2 < M_Z$, or 
$\phino\to\gamma\hino{b}$, for $M_2 > M_Z$, benefits from a strong dynamical 
suppression of the tree-level decays. When $M_2$ gets larger than 1.37$M_Z$, 
$m_{H-Z}^{(-)}$ becomes the next-to-lightest neutralino, while $\hino{b}$ 
is the lightest one. Once more, no dynamical enhancement can occur. 
This rather complex behavior generates sharp and well outlined
contours for the BR(\nrad) in the $(M_1, M_2)$ plane, in the vicinity 
of the diagonal $M_1 = M_2$ (see Sec.~5).

Of course, when $(M_1 - M_2)$ and $(\tgb - 1)$ go away from 0, all the
arguments given above, including the condition (\ref{cond3}), have to 
be intended in a weaker sense. Large radiative BR's can be obtained, even 
in points of the parameter space where $|M_1 - M_2| \ltap M_Z/2$ and 
$\tgb \ltap 2$. As for the validity of the condition (\ref{cond3}) in a 
less restricted case, one has to note that a massless (or very light) state 
is present whenever the equation
\beq 
M_1 M_2 \mu = (M_1 \cos^2 \theta_W + M_2 \sin^2 \theta_W) M_Z^2 
\sin 2\beta
\label{HZ-massless}
\eeq 
approximately holds \cite{Bartl-Matr}. This implies a positive $\mu$. 
For general values of $M_{1,2}$ and $\tgb$, this state is a superposition 
of all the four interaction eigenstates. When $(M_1 - M_2) \to 0$  
its photino component tends to vanish, while its $\hino{b}$ component 
gets smaller and smaller as $\tgb \to 1$.
Hence, in general, in a wide region in the vicinity of the curve 
defined by Eq.(\ref{HZ-massless}) in the plane ($M_1,M_2$),  
the lightest neutralino (or the next-to-lightest one) is a mixed state 
made of Higgsino--Z-ino and no dynamical-enhancement mechanism can be present,
even if $M_1 \simeq M_2$ and $\tgb \simeq 1$. 

In our numerical analysis, we have not found any
case of sizeable dynamical suppression for positive values of $\mu$, 
in the allowed regions. Note that, for a positive $\mu$, a very light 
chargino can be present. In particular, if 
\beq 
M_2 \mu = M_W^2 \sin 2\beta
\label{C-massless} 
\eeq 
the light chargino is massless. This gives rise to ``forbidden'' regions in
the \susy-parameter space in the positive-$\mu$ case, 
where our basic assumption LSP$=\n{1}$ may not be satisfied. \\  
On the other hand, even if $\mc{1} > \mn{1}$, the chargino often turns 
out to be lighter than the $\n{2}$, opening the cascade channels, 
which do not suffer from dynamical suppression.
This happens, for low $\tgb$, only rarely 
for $\mu<0$ and often for $\mu>0$, and provides a further
explanation for the lack of significant BR(\nrad)-enhancement
regions in the positive-$\mu$ case. 

In order to get a general insight of the neutralino physical composition 
pattern, we show in Figs.~2-4 the behavior in the ($M_1,M_2$) plane 
of the quantities 
$A=\langle\protect\n{1}|\protect\phino\rangle^2
    \langle\protect\n{2}|\protect\hino{b}\rangle^2$ \ and \  
$B=\langle\protect\n{1}|\protect\hino{b}\rangle^2
    \langle\protect\n{2}|\protect\protect\phino\rangle^2$ \ \ 
($A,B \le 1$), which express the neutralino physical ``purity'' in 
the dynamical enhancement framework. 
We also study the effect of varying both the $\mu$ and $\tgb$ values in 
the interesting ranges. In the following, we will see that, in order to 
achieve an appreciable dynamical enhancement of BR(\nrad), either A or B 
should be as high as 0.8-0.9. This condition is fulfilled 
in a substantial portion of the ($M_1,M_2$) plane, for low $\tgb$.  
As expected, considerably high values\footnote{Note that in the 
``democratic" case $\langle\protect\n{i}|\protect\phino\rangle^2 = 
\langle\protect\n{i}|\protect\hino{b}\rangle^2 = 1/4$;  
$i = 1, 2$, one would have A=B=1/16 only.} of A, B can generally be achieved 
for $M_1 \simeq M_2$, when $\tgb$ is close to 1. Some deviations from 
the expected behavior in the limit $M_1 = M_2$ and $\tgb = 1$ we
treated above are due to our choice $\tgb = 1.2$. 
In general, the presence of contour lines that delimit an abrupt change of
regime in either A or B generally corresponds to a crossing in the mass
ordering of a physically ``pure" neutralino and a ``mixed" state (or 
of two ``pure'' states). 

\begin{figure}[h]
\centerline{ 
\epsfxsize=0.55\textwidth 
\epsffile{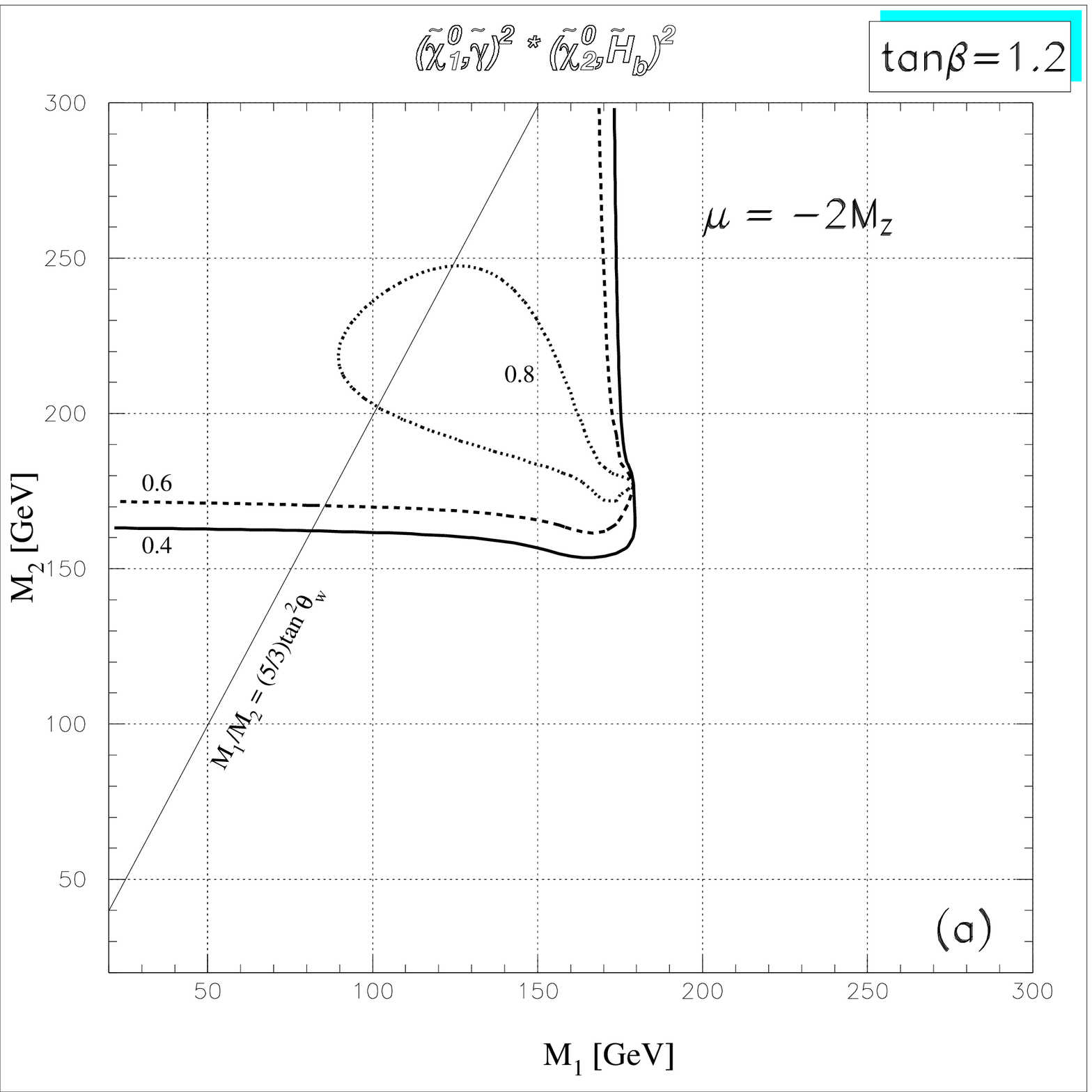}
\hspace{0.5cm} 
\epsfxsize=0.55\textwidth 
\epsffile{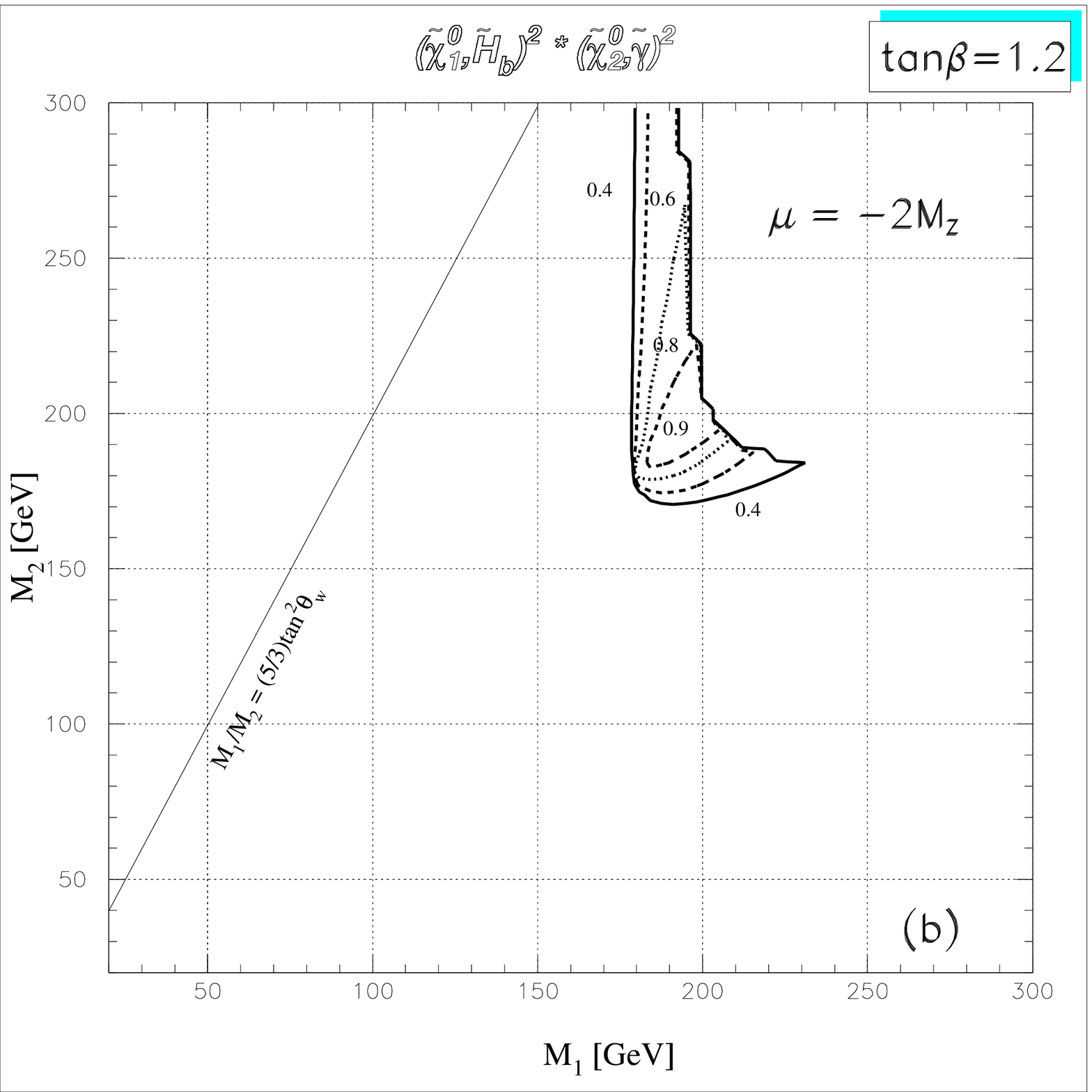}
} 
\caption{Contour plot for the quantities: 
(a) $A=\langle\protect\n{1}|\protect\phino\rangle^2
    \langle\protect\n{2}|\protect\hino{b}\rangle^2$ \ and \  
(b) $B=\langle\protect\n{1}|\protect\hino{b}\rangle^2
    \langle\protect\n{2}|\protect\protect\phino\rangle^2$ \ \ 
($A,B \le 1$), \ \ 
in the case $\tgb = 1.2$, $\mu = -2M_Z$.   
A and B give a hint of the ``purity'' of the limit 
 $\protect\n{1}=\protect\phino$, $\protect\n{2}=\protect\hino{b}$ 
or vice versa. The line corresponding to gaugino-mass unification 
is also shown.} 
\label{NeuComp_mu-2Mz_tgb1p2}
\end{figure} 

For instance, in Fig.~2, proceeding along the $M_1 = M_2$ diagonal 
from small to large $M_{1,2}$ values, one can single out four different
regimes and this can be explained by the discussion above.  
The behavior of A and B along this diagonal is of particular 
interest, as anticipated. Indeed, the latter is the only region where A,
B can substantially exceed the 0.8 level and the dynamical enhancement 
mechanism can be fully effective. 
When $M_{1,2} \ltap 150$ GeV, the lightest neutralino is an almost pure 
photino with mass close to $M_{1,2}$ and the next-to-lightest neutralino is
a mixed Higgsino-Zino state with mass close to 
Eq.(\ref{mhz})$_+$ (the $+$ refers to the sign considered in the equation). 
Also, $\n{3} \approx \hino{b}$ with 
mass close to $-\mu$, and $\n{4}$ is the other mixed state with mass 
close to the absolute value of Eq.(\ref{mhz})$_-$. 
For 150 GeV$\ltap M_{1,2} \ltap -\mu = 2M_Z$ the mostly-$\hino{b}$
and the lighter mixed states exchange their role, becoming the $\n{2}$ 
and the $\n{3}$, respectively. This arrangement is then suitable for 
a dynamical BR(\nrad) enhancement with $\n{1} \simeq \phino$ and 
$\n{2} \simeq \hino{b}$ [cf. Fig.~2(a)]. 
For $2M_Z \ltap M_{1,2} \ltap 200$ GeV, the mass ordering of the dominantly 
$\phino$ and $\hino{b}$ states is exchanged, but one still has a 
scenario with dynamical enhancement [cf. Fig.~2(b)]. Note also 
that the two mixed states exchange their role as well, the one 
corresponding to the negative mass eigenvalue becoming lighter than the 
other. This double level crossing takes actually place in the highly 
degenerate point $M_1 = M_2 = -\mu \sindb \simeq 179.4$ GeV, where 
all the contour levels in Fig.~2 tend to crowd. Points of this kind 
will turn out to be of relevance for the kinematical mechanism too 
(cf. Sec.~4). 
Finally, when $M_{1,2} \gtap 200$ GeV the negative eigenvalue
corresponding to the mixed state crosses the mass level ($\simeq M_{1,2}$)
of the almost pure photino, which then becomes the $\n{3}$. Then, the 
dynamical mechanism stops working, in spite of the presence of an 
almost pure $\hino{b}$ as $\n{1}$. 

\begin{figure}[h]
\centerline{ 
\epsfxsize=0.55\textwidth 
\epsffile{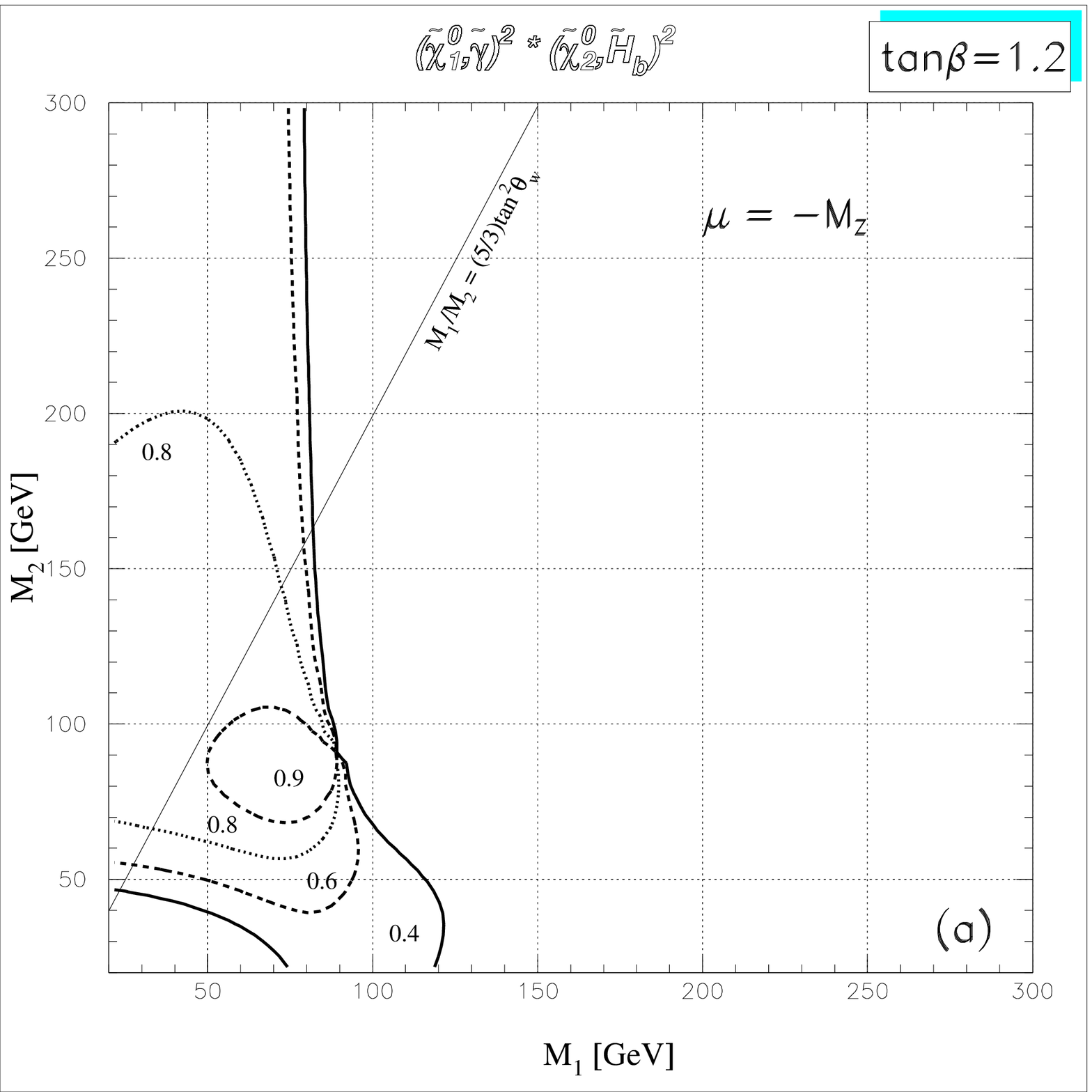}
\hspace{0.5cm} 
\epsfxsize=0.55\textwidth 
\epsffile{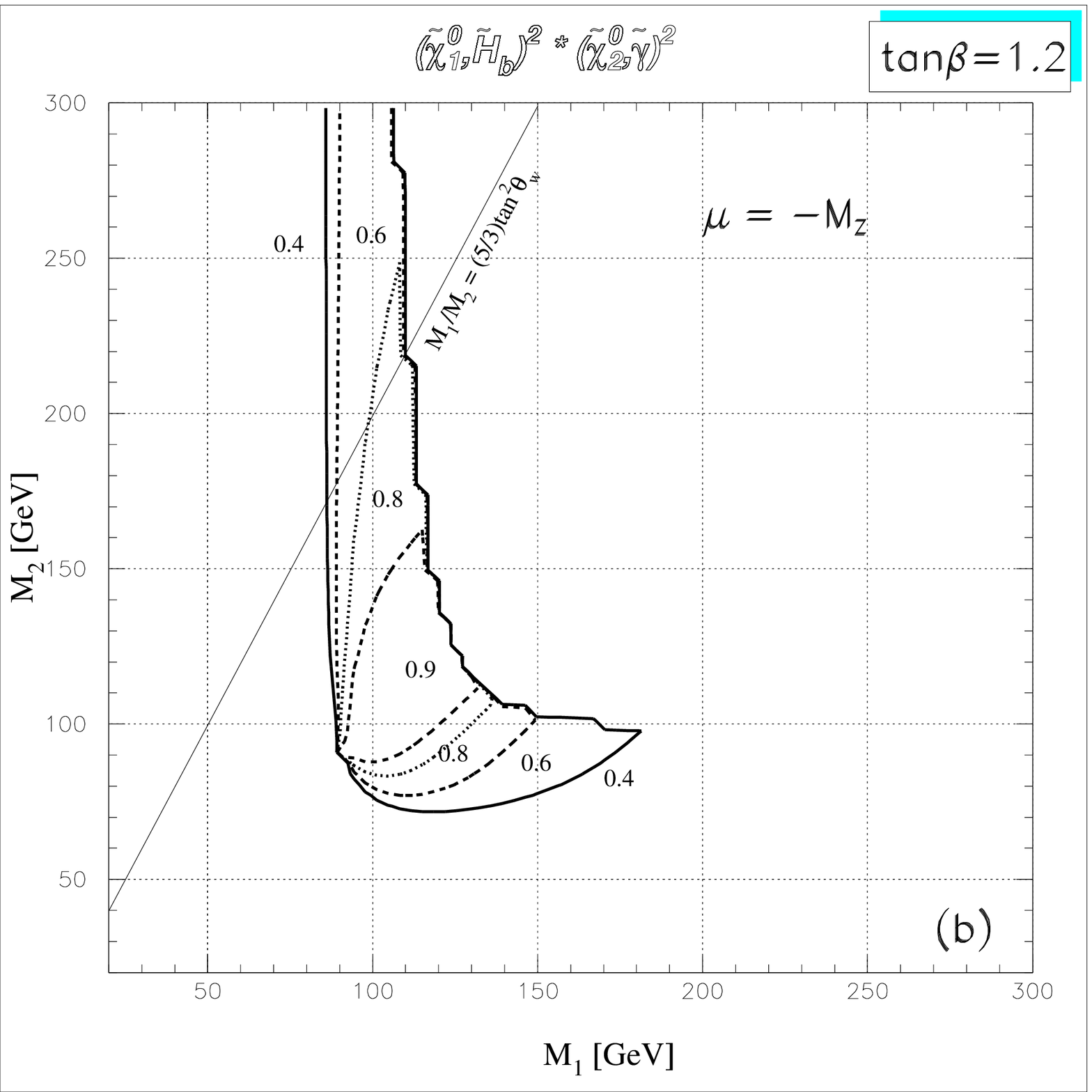}
}
\caption{The same as in Fig.\protect\ref{NeuComp_mu-2Mz_tgb1p2}, but 
for $\mu=-M_Z$.} 
\label{NeuComp_mu-Mz-tgb1p2}
\end{figure} 

In Fig.~3, we show how the general picture for A and B evolves 
when $\mu$ goes from $-2M_Z$ to $-M_Z$.  
One can check that the situation is qualitatively similar to the previous 
case, once the whole structure of the contour plots in the 
($M_1, M_2$) plane is shifted toward the new crossing point 
$M_1 = M_2 = -\mu \sindb \simeq 89.7$ GeV. Note, however, that here 
the quantity A can reach higher values (of order 0.9 or more),
and the region where B is large is also wider in Fig.~3 with respect to
Fig.~2. Furthermore, it is interesting that it is possible to approach 
high values of A and B ($A \simeq 0.9$ and $B \simeq 0.85$ in a certain 
$M_{1,2}$ interval) even in the gaugino-mass unification case. 

For $\tgb$ as high as 4 (Fig.~4), A and B never reach 0.8 and, 
consequently, never prompt a sufficient dynamical BR(\nrad) enhancement. 

\begin{figure}[h]
\centerline{ 
\epsfxsize=0.55\textwidth 
\epsffile{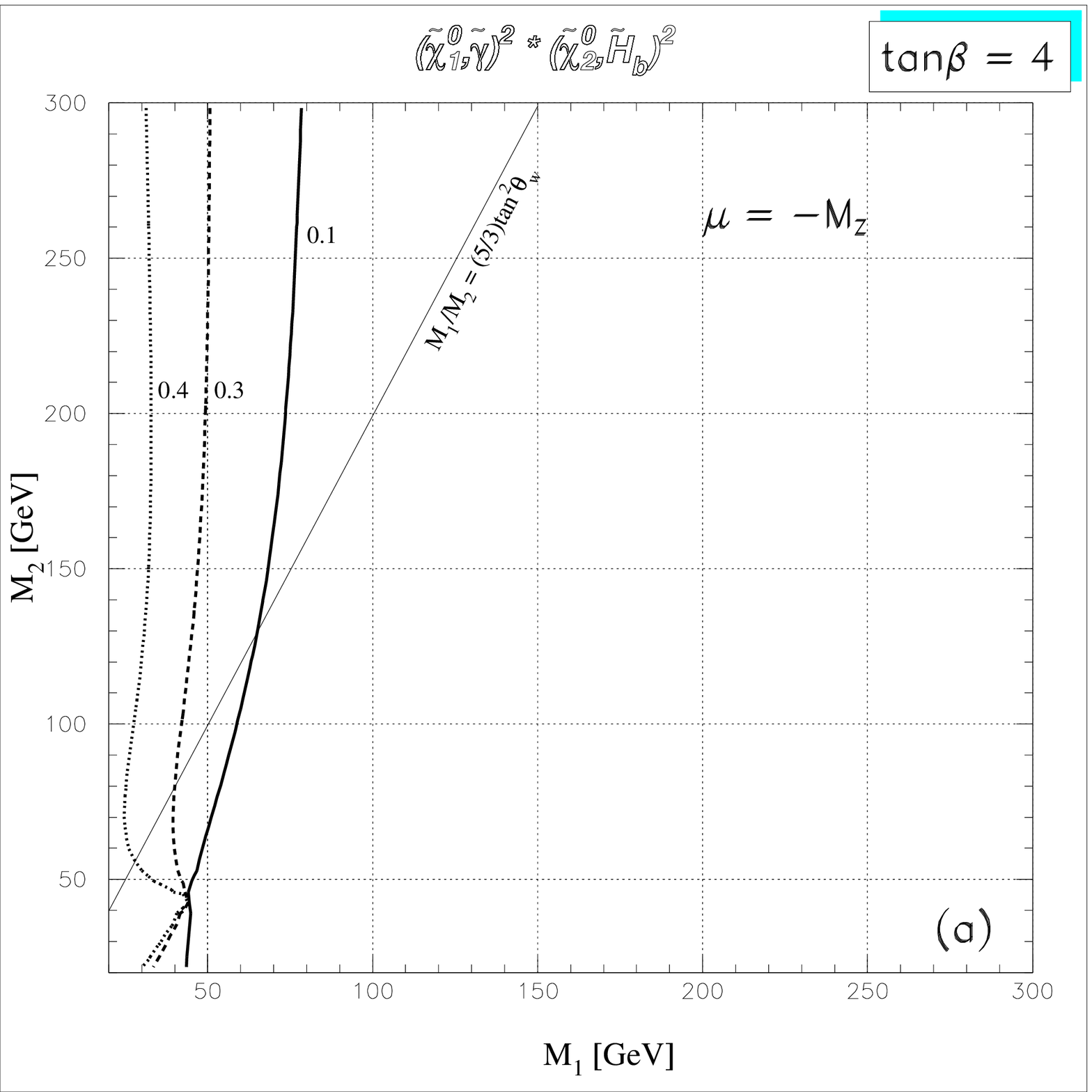}
\hspace{0.5cm} 
\epsfxsize=0.55\textwidth 
\epsffile{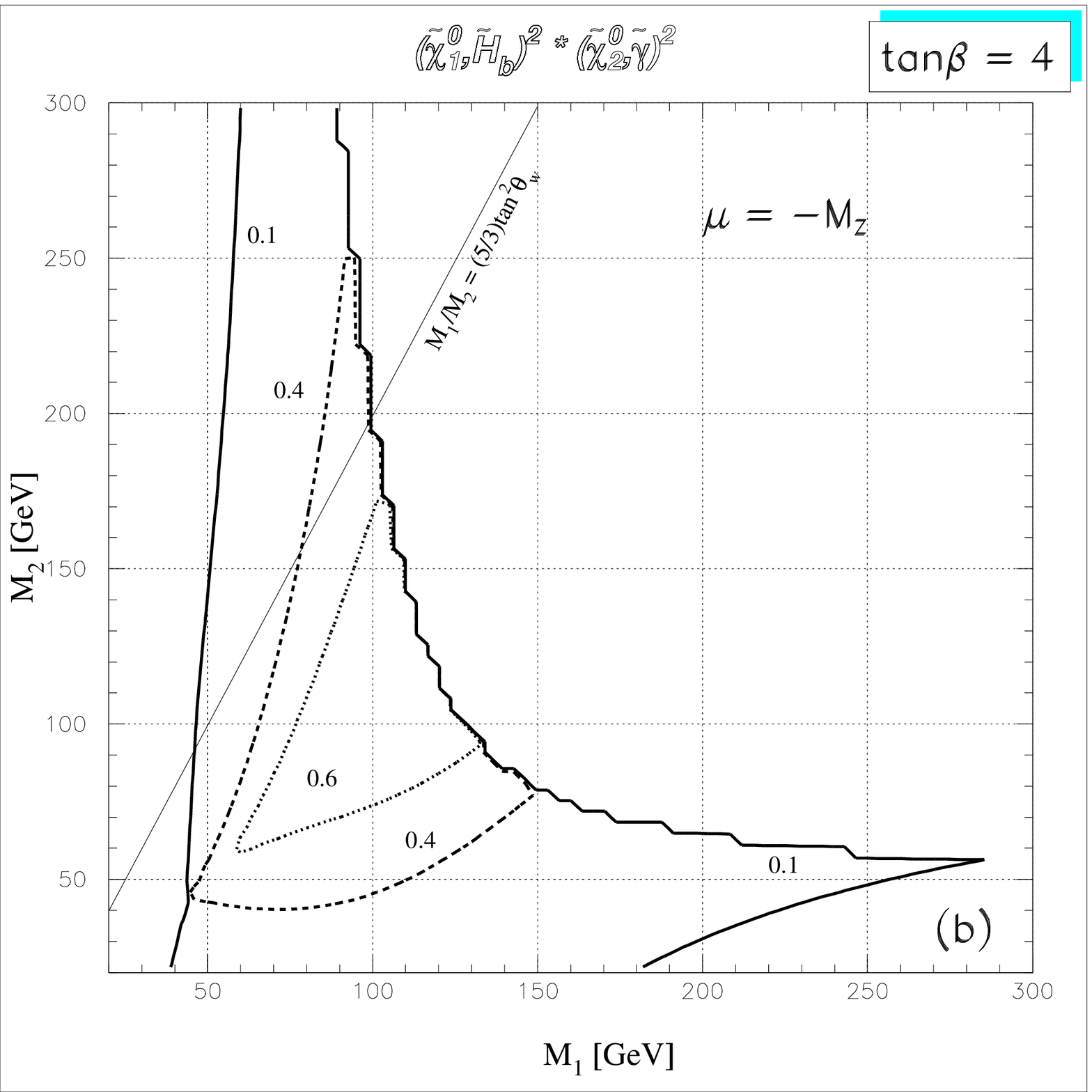}
}
\caption{The same as in Fig.\protect\ref{NeuComp_mu-2Mz_tgb1p2}, but 
for $\mu=-M_Z$ and $\tgb = 4$.} 
\label{NeuComp-mu-Mz-tgb4}
\end{figure} 

\section{Kinematical \protect\nrad\ enhancement} 
\noindent
As anticipated in Sec.~2, when $\n{1}$ and $\n{2}$ are almost degenerate
and ($\mn{1} - \mn{2}$) is smaller than about 10 GeV, for $\mn{1,2} \approx 
M_Z$, the radiative decay is enhanced by a large, purely kinematical
factor. The latter can actually turn out to be an overall factor, especially 
when the contribution of the sfermion exchange to the $\n{2}$ tree-level 
decays is suppressed. Hence, one has an optimization of this ``kinematical''
enhancement for heavy scalar masses and/or small gaugino 
components in $\n{1}$ and/or $\n{2}$, as can be inferred by comparing 
Eqs.(\ref{Gamma_Z}) and (\ref{Gamma_sf}). Indeed, after summing over flavors 
and colors in Eq.(\ref{Gamma_sf}), if (some of) the $C^{L,R}_{\tilde{A}}$
are not far from 1 and/or (some of) the $\tilde{f}_{L,R}$ are not much heavier
than about $M_Z$, the sfermion-exchange channels 
and the interferences can dominate in this regime.  
The net effect of this can be to slow down or even prevent the introduction 
of an effective kinematical enhancement, so that BR(\nrad) might get large, 
if ever, only for neutralino mass differences smaller than a few GeV, 
which we know to be a potentially dangerous region to explore, at least 
in our approximations. 

Given the above considerations, the next step is to find out where in 
the \susy-parameter space the masses of the two lightest neutralinos are 
almost degenerate. To this aim, one has to consider the 
4$^{\rm th}$-degree eigenvalue equation associated with the neutralino 
mass matrix~(\ref{Neu-Matrix}) 
\newpage 
\bea
0 & = & m_i^4 + A m_i^3 + B m_i^2 + C m_i + D  \nonumber \\ 
  &   & {\rm with:}  \nonumber   \\ 
A & = & -{\rm Tr}({\cal M}_{\n{}}) = - (M_1 + M_2)          \; , \nonumber \\
B & = & M_1 M_2 - \mu^2 - M_Z^2                             \; , \nonumber \\ 
C & = & \mu^2(M_1+M_2)+M_z^2(M_1\cthwq+M_2\sthwq-\mu\sindb) \; , \nonumber \\ 
D & = & {\rm det}({\cal M}_{\n{}}) = 
\mu[M_Z^2(M_1\cthwq+M_2\sthwq)\sindb - \mu M_1 M_2]    \; , \label{Neu-eq}  
\eea 
where $m_i$ ($i = 1, \ldots 4$) is the general neutralino mass eigenvalue. 
Then, one has to force Eq.(\ref{Neu-eq}) (which, arising from an Hermitian, 
real and symmetric matrix, has four real roots) to have (at least) either two 
identical roots or two opposite roots, $m_i^o = \pm m_j^o$ (when using  
the superscript $^o$, we generally refer to a degenerate 
eigenvalue).\footnote{The physical neutralino mass is given 
by $\mn{i} = |m_i|$.} 

Exact expressions for the neutralino masses and mixing can be found in 
Ref.\cite{ElkBarger}. Here, we are mainly interested in special cases for 
which it is possible to extract approximate formulas, more useful for a  
physical interpretation. 

Involving a linear combination of 12-dimensional terms 
(where up to the 4$^{\rm th}$ power of one of the coefficients A, B, C, D 
can appear), the general necessary condition to get two identical roots from 
Eq.(\ref{Neu-eq}), seems far too complicated to give any useful information
and even to be displayed here. As for the case of two opposite roots, 
a simple necessary condition in terms of A, B, C, D can be 
derived: 
\beq 
A^2 D + A B C + C^2 = 0 \; . 
\label{Opp-Cond}
\eeq 
Unfortunately, Eq.(\ref{Opp-Cond}), when translated in terms of $M_1$, 
$M_2$, $\mu$, $\sindb$ and $\sthwq$, turns out to be quite complex too. 
Thus, in the following, we consider only interesting limits of the \susy\ 
parameters, such as $\tgb \to 1$ or $M_1 \to M_2$. 
In this way, by reducing to zero the off-diagonal terms in one or both of 
the $2 \times 2$ blocks of the matrix (\ref{Neu-Matrix}), one considerably 
simplifies the eigenvalue equation (\ref{Neu-eq}) and allows to disentangle 
the relevant degeneracy scenarios. In this way, we will single out some 
{\it sufficient} conditions, which ensure exact or approximate degeneracy 
between the two lightest neutralino states. 
This procedure will be supported by an extensive numerical study 
(scanning the whole \susy-parameter space) of the relevant mass 
splitting ($\mn{2}-\mn{1}$), which is shown in the following, 
as well as a numerical analysis of BR(\nrad), that we present in the next 
section. On the basis of the numerical analysis, we found that {\it necessary} 
conditions for an exact degeneracy of the two lightest neutralino states
are
\beq 
\tgb = 1 \; \; \; {\rm or} \; \; \; M_1 = M_2 \; , 
\label{nec-cond} 
\eeq
with the additional requirement $\mu < 0$. 
The latter directly translates in some {\it necessary} conditions 
for a sizeable kinematical enhancement of BR(\nrad).
Interestingly enough, these conditions are the same as the ones we 
found in Sec.~3 for a dynamical enhancement, although 
in the latter case, the two requirements on $\tgb$ and $M_{1,2}$ 
have to be fulfilled at the same time. 
The conditions (\ref{nec-cond}) for a BR(\nrad) kinematical enhancement 
are of course interesting only in regions allowed by LEP1-1.5 data. 
Furthermore, we do not consider here asymptotic regimes with 
$M_{1,2} \gg M_Z$ and/or $|\mu| \gg M_Z$, or $M_{1,2} \ll M_Z$ and/or 
$|\mu| \ll M_Z$. Indeed,  for instance, the limit 
$|\mu| \ll M_{1,2}, \; M_Z$ leads to two almost degenerate Higgsinos-like 
neutralinos, as discussed above and in Ref.\cite{Haber-Wyler}. 
Other asymptotic cases with possible degeneracy have been 
mentioned in Sec.~2. Here, we will limit our analytical and numerical 
analyses to the region where $|\mu|$ and $M_{1,2}$ fall both in the interval 
$[M_Z/4, 4 M_Z]$.  

We stress that here we are interested to the degeneracy of the two lightest 
neutralinos. This singles out only a few among the possible degeneracy 
scenarios for the four neutralinos, and makes the analysis more involved.   

In order to find {\it sufficient} conditions for the degeneracy of the two 
lightest neutralinos, let's now first consider the limit $M_1 = M_2$. 
Then, it is convenient to solve the equation (\ref{Neu-eq}) 
just in terms of $M_2$ as a function of the generic eigenvalue $m_i$. 
For $m_i \neq \pm \mu$, one then gets two branches
\bea
(M_2)_+  & = & m_i  \; ,              \nonumber \\
(M_2)_-  & = & m_i - M_Z^2\frac{m_i + \mu \sindb}{m_i^2 - \mu^2} \; . 
\label{solve-m2}   
\eea 
Note that the branch $(M_2)_+$ describes the behavior of one neutralino 
mass eigenvalue only, while the branch $(M_2)_-$ is threefold and corresponds 
by itself to three generally different eigenvalues.
The two main branches $(M_2)_{\pm}$ intersect in a point (corresponding to 
a degeneracy with same-sign eigenvalues: $m_{i,j}^o = M_2$), whenever 
\beq  
M_1 = M_2 = -\mu \sindb   \; .    \label{deg-cond-alfuno} 
\eeq 
This case is quite interesting. Indeed, one has: 
\bea
 m_1^o = m_2^o  & = & M_1 = M_2 = -\mu \sindb \; , \label{deg-sol-alfuno} \\ 
-m_3^o = m_4^o  & = & \sqrt{\mu^2+M_Z^2}  \; ,     \label{deg-sol-spur}  
\eea 
that is, the two mass eigenvalues with the lower absolute values coincide, 
while the two eigenvalues with the higher absolute values are opposite. 
As for the composition, one of the two light degenerate states
is a pure photino, while the other is a mixture of $\Zino$ and 
$\hino{b}$, with $\langle\Zino|\n{2}\rangle^2 = 
\mu\cosdb/\sqrt{M_Z^2+\mu^2\cosdb}$. 
Thus, the scenario of Eq.(\ref{deg-cond-alfuno}) is relevant both for 
the kinematical and for the {\it reduced} dynamical enhancement
(cf. Sec.~2), since the $\Z$-exchange contribution to the tree-level 
decays is highly suppressed. The degeneracy corresponding to 
Eq.(\ref{deg-sol-alfuno}) is of course removed when $M_1$ and $M_2$ get  
far apart. Nevertheless, one can check that for $M_1 - M_2 \sim 10$ GeV 
the degeneracy may still be effective for a sizeable 
kinematical enhancement of the radiative decay. 

In principle, one can get other degeneracy scenarios by using 
Eqs.(\ref{solve-m2}) in different ways, but we will show in the following 
that they are not relevant for our purposes. First of all, it is clear that 
the degeneracy between the heavier neutralinos corresponding to
Eq.(\ref{deg-sol-spur}) cannot be a direct result of the intersection of 
the two main branches $(M_2)_{\pm}$. 
However, it can happen that two of the three sub-branches of the branch 
$(M_2)_-$, for a given value of $M_2$, correspond to two opposite 
eigenvalues, and hence to degenerate neutralino masses. This is the
case when $(m_i^2-\mu^2) = M_Z^2$, which 
is nothing but the degeneracy (\ref{deg-sol-spur}). The additional fact that 
in such a scenario one necessarily has also $M_2 = -\mu \sindb$ [corresponding 
to a {\it real} intersection of the two main branches $(M_2)_{\pm}$ and to the
further degeneracy (\ref{deg-sol-alfuno}) of two always lighter neutralino 
states] makes the above circumstance negligible for us here. 
One can look for other cases of degeneracy by considering the
possibility that one of the eigenvalues described by the branch $(M_2)_-$
has the same absolute value of the one corresponding to the branch $(M_2)_+$, 
but opposite sign. This corresponds to solving the equation 
\beq 
M_Z^2 \frac{M_2 - \mu \sindb}{M_2^2 - \mu^2} - 2M_2 = 0 \; , 
\label{deg-cond-alfuno-opp} 
\eeq 
while, for the two degenerate eigenvalues, one has: 
$m_i^o = - m_j^o = M_1 = M_2$. The solutions of 
Eq.(\ref{deg-cond-alfuno-opp}) are in general complicated expressions, but 
one can easily find them numerically. 
After such an analysis, we did not encounter any further case of
exact neutralino mass degeneracy from Eq.(\ref{deg-cond-alfuno-opp}) 
in the limit $(M_1 - M_2) \to 0$ relevant for the kinematical
enhancement, in regions allowed by the LEP data.  

Relaxing the limit $M_1 = M_2$, a different {\it necessary} condition to
get mass degeneracy in Eq.(\ref{Neu-eq}) is, indeed, $\tgb \to 1$. In this 
limit, contrary to the previous case, the easiest part of the neutralino
mass matrix is the Higgsino sector. 
Then, we can solve the eigenvalue equation with respect
to $\mu$ as a function of $m_i$, in order to get other {\it sufficient} 
conditions, and scenarios of interest for the kinematical enhancement. 
One then finds again two branches
\bea 
(\mu)_-  & = & -m_i  \; , \nonumber \\ 
(\mu)_+  & = & m_i - M_Z^2\frac{m_i - M_1\cthwq - M_2\sthwq}
               {(m_i-M_1)(m_i-M_2)}  \; , \label{solve-mu} 
\eea 
where, similarly to the case of Eqs.(\ref{solve-m2}), the branch $(\mu)_-$ 
describes a single neutralino mass eigenvalue and the branch $(\mu)_+$
corresponds to three different eigenvalues. 
Note that here, contrary (and complementary) to the case of 
Eq.(\ref{solve-m2}), one can have $m_i = \pm \mu$, but not $m_i = M_1$
or $M_2$. 
By using Eqs.(\ref{solve-mu}), one can single out degeneracy scenarios 
with $m_i^o = \pm m_j^o = \mu$.
In order to realize the case with two same-sign degenerate eigenvalues, 
the general condition 
\beq 
M_Z^2\frac{\mu+M_1\cthwq+M_2\sthwq}{(\mu+M_1)(\mu+M_2)} - 2 \mu = 0 \; , 
\label{deg-cond-tgbuno}
\eeq 
must hold. Again, the corresponding explicit solutions are rather complex, 
but one can solve Eq.(\ref{deg-cond-tgbuno}) numerically to find the 
regions of the parameter space interesting for the kinematical enhancement. 
Since, in contrast to the case of Eq.(\ref{deg-cond-alfuno-opp}), 
Eq.(\ref{deg-cond-tgbuno}) gives rise to interesting scenarios, it is useful 
to consider here, in addition to $\tgb \to 1$, the special limit $\sthwq=0$, 
which allows a simplified analytical treatment. Indeed, in this way, two free 
gaugino mass parameters are still present, but the mixing in the gaugino 
sector disappears. Furthermore, this limit is not too far away from the real 
physical case $\sthwq \simeq 0.23$. Equation (\ref{deg-cond-tgbuno}) gives 
then the solutions (when $\tgb = 1$)
\beq 
m_i^o = m_j^o = -\mu = \frac{1}{2} (M_2 \mp \sqrt{M_2^2 +
2M_Z^2}) \;  .                              \label{deg-sol-AB}
\eeq 
The existence of the solutions (\ref{deg-sol-AB}) (not their exact 
form), is independent of the limit $\sthwq \to 0$ and the 
corresponding exact degeneracy is removed only for $\tgb \neq 1$. 
The introduction of this limit allows us to disentangle 
other two interesting cases, as long as $\tgb = 1$, that is
\beq 
m_i^o = m_j^o = - \mu = M_1 \; , \label{deg-sol-C}
\eeq 
and, when the condition
\beq 
M_1 - \frac{M_Z^2}{M_1 - M_2} - \mu = 0 \label{deg-cond-D}
\eeq 
is fulfilled, 
\beq 
m_i^o = m_j^o = M_1 = \mu-\frac{M_Z^2}{M_1 - M_2} \; . \label{deg-sol-D}
\eeq 
The latter cases, needing $m_i = M_1$, cannot be directly derived from the 
two branches in Eq.(\ref{solve-mu}). The correct procedure 
to get them is to solve the eigenvalue equation with respect to 
$\mu$ or $M_1$, by applying both the limits $\tgb \to 1$ 
and $\sthwq \to 0$ simultaneously. \\ 
In order to understand the nature of these additional solutions and their 
link with the limit $\sthwq \to 0$, some further explanation is needed. 
A solution corresponding to Eq.(\ref{deg-sol-C}) survives when 
$\sthwq \neq 0$, although the expression for the degenerate mass eigenvalues 
receives some corrections, as in the case of solutions (\ref{deg-sol-AB}). 
What makes case (\ref{deg-sol-C}) different from the previous ones is that 
the corresponding degeneracy is not removed when $\tgb$ goes away from 1 and 
both $\sthwq \neq 0$ and $\tgb \neq 1$ are needed to do the job. In this 
sense, the degeneracy corresponding to Eq.(\ref{deg-sol-C}) is more 
``solid'' than the others. As for solution (\ref{deg-sol-D}), instead, it 
represents a spurious case which does not correspond anymore to an exact
degeneracy when $\sthwq \neq 0$, but only to a case where two neutralino 
mass eigenvalues are close to each other (in the limit $\tgb \to 1$), 
although not quite equal. To get an effective kinematical BR(\nrad) 
enhancement, we are interested in scenarios where neutralino mass differences 
of order 10 GeV or less arise. In order to obtain such a small 
($m_i^o - m_j^o$) when Eq.(\ref{deg-sol-D}) holds, one needs 
$M_1, M_2 \ll M_Z, |\mu|$, with $|\mu| \approx$ TeV [because of condition 
(\ref{deg-cond-D})]. Then, the quantity ($M_1 \cthwq + M_2 \sthwq$) in 
Eq.(\ref{Neu-eq}) can be treated in the same way as it would be in the limit 
$\sthwq \to 0$. Therefore, we will neglect this possibility (corresponding 
to an asymptotic case, already excluded by LEP) and we will focus on 
solutions (\ref{deg-sol-AB}), (\ref{deg-sol-C}). 

We stress that the existence of the solutions (\ref{deg-sol-AB}), 
(\ref{deg-sol-C}) (not the exact expression for the degenerate eigenvalues) 
is independent of the limit $\sthwq \to 0$ and the corresponding exact 
degeneracy is removed only for $\tgb \neq 1$. Anyway, the simplified 
solutions we found allow us to emphasize some remarkable properties which 
remain valid with a good approximation for $\sthwq \simeq 0.23$ (and, often, 
even for $\tgb \simeq 1$, rather than exactly 1). For instance, the solutions 
corresponding to (\ref{deg-sol-AB})$_+$ and (\ref{deg-sol-C}) are only
possible for negative values of $\mu$ [even $\mu < -M_Z/\sqrt{2}$ in the 
case (\ref{deg-sol-AB})$_+$)].\footnote{The subscript ${\pm}$ of the equation 
number picks out one of the two possible signs in Eq.(\ref{deg-sol-AB}).}
In contrast, the solution (\ref{deg-sol-AB})$_-$ is allowed only for 
$0 \le \mu \le M_Z/\sqrt{2}$ (that is in a region that, particularly 
for small $\tgb$, is excluded by LEP1-1.5 data, due to the
presence of a light chargino). Hence, solution (\ref{deg-sol-AB})$_-$ 
will not play an important role in the following discussions. 
Also, note that the solution (\ref{deg-sol-C}) is present irrespective of the
particular value of $M_2$ and, similarly, the solutions (\ref{deg-sol-AB})  
do not depend on $M_1$, in the limit $\tgb \to 1$, $\sthwq \to 0$ and 
exact degeneracy. For $\sthwq \simeq 0.23$, the solution (\ref{deg-sol-AB}) 
[(\ref{deg-sol-C})] develops a weak dependence on $M_1$ [$M_2$], as 
will be shown in the following numerical study. 

As for the limit $\tgb \to 1$, up to now we only took care of deriving
some sufficient conditions for the degeneracy of {\it any pair} of
neutralinos. Now, we need to check when the degeneracy scenarios we singled 
out actually concern the two lightest neutralino mass eigenstates. We will 
focus on the more interesting scenarios (\ref{deg-sol-AB})$_+$ and 
(\ref{deg-sol-C}). Which pair of mass eigenstates is involved in the 
degeneracy depends also on the parameters not directly entering the 
approximate conditions (\ref{deg-sol-AB}), (\ref{deg-sol-C}), in a generally 
simple way. For a given value of $|\mu|$, typically one observes that, 
for $M_1 \; [M_2] > |\mu|$, the degeneracy of the kind (\ref{deg-sol-AB})$_+$ 
[(\ref{deg-sol-C})] indeed concerns the two lightest neutralino states. 
On the other hand, as long as $M_1 \; [M_2] < |\mu|$, the solution for the 
mass degeneracy corresponds to $\mn{2} = \mn{3} = -\mu$ and, thus, does not 
give rise to any kinematical suppression.  

All the $\tgb = 1$ scenarios above are derived by forcing the two branches in 
Eqs.(\ref{solve-mu}) to meet in a point of the ($m_i, \mu$) plane, 
corresponding to a {\it strict} degeneracy $m_i^o = m_j^o$. 
However, as done above in the limit $(M_1 - M_2) \to 0$ with 
Eq.(\ref{deg-cond-alfuno-opp}), here too we must take into account the 
additional possibility of a degeneracy with opposite eigenvalues 
$m_i^o = -m_j^o$. Imposing the corresponding condition $(\mu)_+ = -(\mu)_-$ 
in Eqs.(\ref{solve-mu}) leads to the interesting case 
\beq 
m_i^o = -m_j^o = \mu = M_1 \cthwq + M_2 \sthwq \; ,  
\label{deg-nosign} 
\eeq 
which satisfies the Eq.(\ref{Opp-Cond}) and is only realized for positive 
values of $\mu$. 
This gives rise to a scenario where a pure $\hino{b}$ is degenerate  
with a superposition of the other interaction eigenstates. 
The other two neutralinos are mixed $\phino$-$\Zino$-$\hino{a}$ states too, 
and correspond to the mass eigenvalues
\beq 
m_{\phino-\Zino-\hino{a}}^{(\pm)} = 
\frac{1}{2} \left[ M_1 + M_2 \pm \sqrt{(M_1 - M_2)^2 + 4 M_Z^2} \right] \; .
\label{mhzf} 
\eeq 
On the other hand, the degeneracy must involve the two lightest states. 
It is easy to show with analytical arguments that this can never happen 
in regions allowed by LEP1-1.5 data. Hence, the scenario of Eq.(\ref{mhzf}) 
is not relevant for the radiative neutralino decay. 

Finally, it is interesting to note that in the limit where both 
$\tan\beta \to 1$ and $(M_1 - M_2) \to 0$, both the kinematical and 
the dynamical enhancements can be optimized at the same time in the 
special point $M_1 (= M_2) = -\mu$, where the two lightest neutralinos
are always a pure $\phino$ and a pure $\hino{b}$    
[cf. Sec.~3, Eqs.(\ref{mhz}) and Eqs.(\ref{deg-sol-alfuno}), 
(\ref{deg-sol-spur})]. We will see, in the numerical analysis of Sec.~5, 
that a considerably wide region of the \susy-parameter space where high 
BR(\nrad) values are realized is centered on this highly degenerate point.

As for the kinematical enhancement, by using numerical methods, we did
not find any other clear case of $\n{1}$-$\n{2}$ exact mass degeneracy,
besides the ones we have described above. Also, it was not possible 
to achieve an approximate degeneracy (at the level of a 10 GeV mass 
difference) either, in regions of the \susy\ space where the {\it necessary} 
conditions (\ref{nec-cond}) are rather far from being valid. 
This does not mean we listed all the possibilities for
neutralino mass degeneracy. For instance, one can consider 
the case $D={\rm det}({\cal M}_{\n{}})=0$. This can be achieved either when 
$\mu = 0$ and/or $M_{1,2} = 0$, or whenever Eq.(\ref{HZ-massless}) holds. 
The first option was already considered among the asymptotic cases.  
In the second case, one is left with a simplified eigenvalue
equation, which gives rise to other degeneracy scenarios.\footnote{ 
For instance, one finds a non-trivial degeneracy, for a given value of 
$\tgb$, in the case 
$M_1 = M_2 = (M_Z/2) \left(1 + \sqrt{1+8 \sin^2 2\beta} \right)^{1/2}$
\ \ \ and \ \ \  
$\mu  = (M_Z/\sqrt{2}) \left(\sqrt{1+8 \sin^2 2\beta} -1 \right)^{1/2}$.
The corresponding degenerate neutralinos are $\n{2}$ and $\n{3}$, with
opposite mass eigenvalues.} 
However, these scenarios always give rise also to at least a null mass
eigenvalue; hence, the degeneracy can only concern the $\n{2}$ and the $\n{3}$ 
or the $\n{3}$ and the $\n{4}$. Other complex degeneracy situations, not of
interest here, can be constructed. 

\vspace{0.3cm}

In summary, the relevant approximate scenarios for the BR(\nrad)
kinematical enhancement are given, for $\mu < 0$, by 
Eq.(\ref{deg-sol-alfuno}), when $M_1 \simeq M_2$, and by 
Eqs.(\ref{deg-sol-AB})$_+$ and (\ref{deg-sol-C}), when $\tgb \simeq 1$.  
In order to get a clear picture of the non-trivial behavior of the neutralino
mass degeneracy, we now show a set of contour plots for the two lightest 
neutralino mass difference ($\mn{2} - \mn{1}$) in the ($M_1,M_2$) plane,
for different values of $\mu$ and $\tgb$. 

\begin{figure}[h]
\centerline{ 
\epsfxsize=0.55\textwidth 
\epsffile{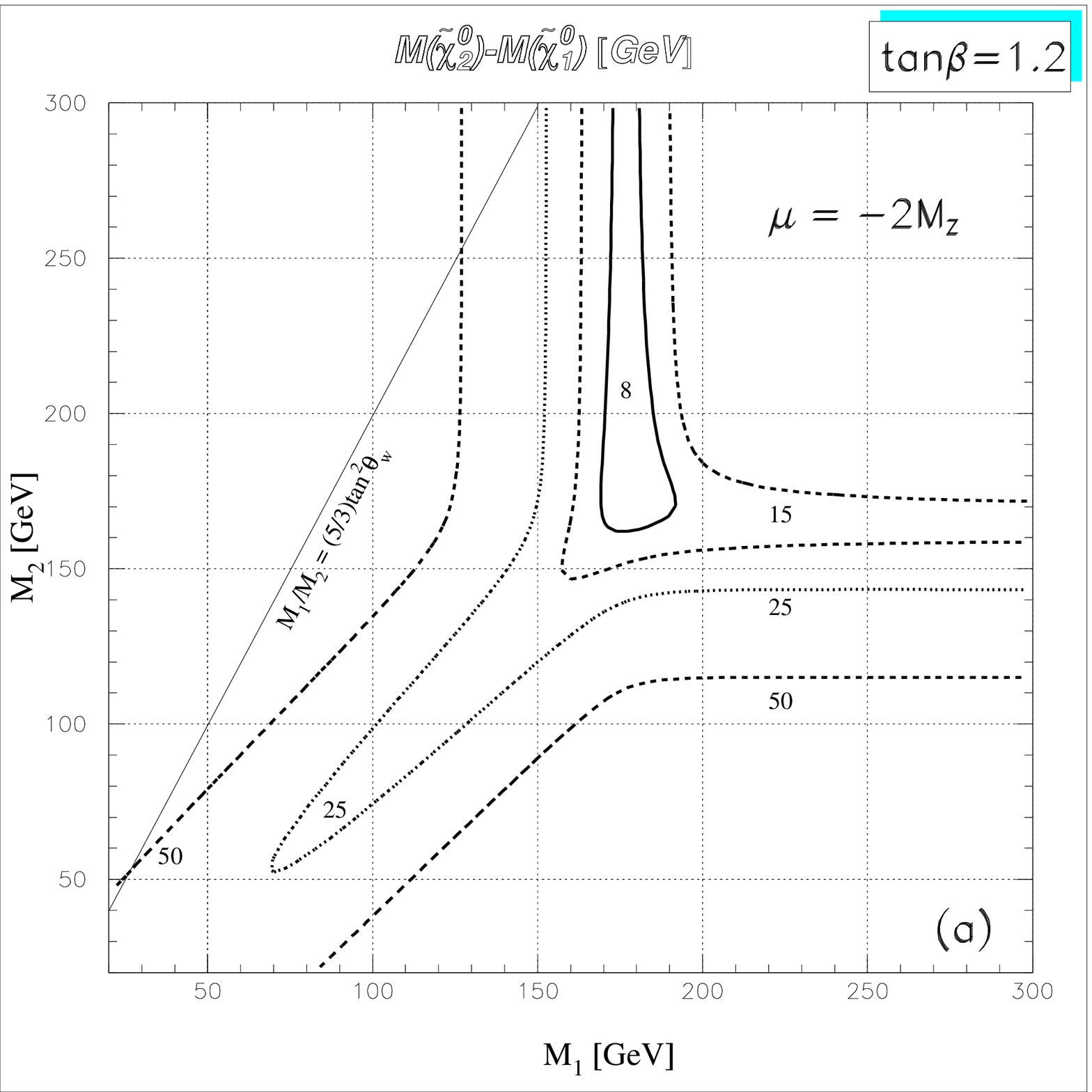}
\hspace{0.5cm} 
\epsfxsize=0.55\textwidth 
\epsffile{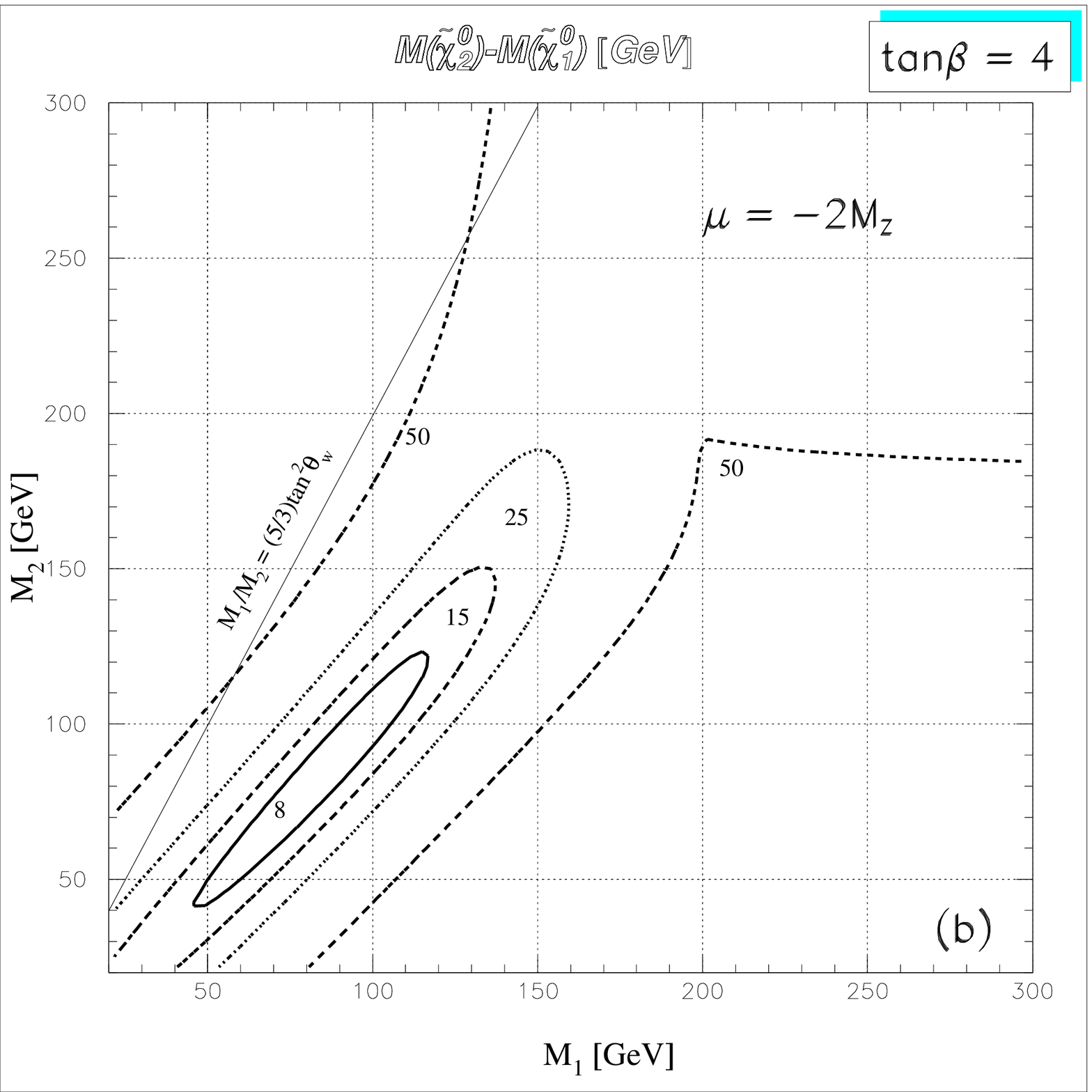}
}
\caption{Contour plot for the difference (in GeV) between the two lightest
neutralino masses in the ($M_1,M_2$) plane for $\mu=-2M_Z$, and 
$\tgb = 1.2$ (a) and 4 (b).  Different levels are represented by lines of
different style. The straight line corresponding to the gaugino mass 
unification is also shown.} 
\label{N2-N1mass_mu-2Mz_tgb1p2}
\end{figure} 

In Fig.\ref{N2-N1mass_mu-2Mz_tgb1p2}, the case $\mu = -2M_Z$ is shown 
for $\tgb = 1.2$ (a) and 4 (b). The line corresponding to the gaugino mass 
unification is also plotted. The general pattern is highly non-trivial 
and quite dependent on $\tgb$. In Fig.\ref{N2-N1mass_mu-2Mz_tgb1p2}(a), 
one can easily note the presence of two quite narrow bands, a vertical one 
and a horizontal one. The horizontal one corresponds to the region where 
the degeneracy scenario of Eq.(\ref{deg-sol-AB})$_+$ is approximatively 
realized for $\tgb$ close to 1 and is well outlined by the 15-GeV 
mass-difference contour. As anticipated, the band is there only for 
$M_1 \gtap -\mu = 2M_Z$ and shows only a weak dependence on $M_1$. Indeed, 
by solving numerically Eq.(\ref{deg-cond-tgbuno}) for the general condition
of degeneracy, we found that the contour line of exact $m_1^o = m_2^o$
degeneracy, in the limit $\tgb=1$, passes through the points 
$(M_1, M_2) = (200, 168.9)$; (250, 166.2); (300, 165.7) GeV, showing a small 
dependence on $M_1$ due to the finite value of $\sthwq$.    
Note that the same equation indicates the presence of a degeneracy also for  
$M_2 =$ 163.7 and 161.5 GeV, respectively for $M_1 =$ 100 and 150 GeV, 
but this corresponds to a case with $m_2^o = m_3^o$, which of course does 
not show on the figure and is not interesting here. 
The vertical band, well outlined by the 8 and 15 GeV contours, represents 
the degeneracy (\ref{deg-sol-C}). Again, the band is present only for 
$M_2 \gtap 2M_Z$ and has only a weak dependence on $M_2$, depending on the 
finite value of $\sthwq$. The contour line of exact degeneracy passes now 
through the points $(M_1, M_2) = (179.7, 200)$; (178.2, 250); (177.7, 300) GeV.
Again, there is also a degeneracy concerning $\n{2}$ and $\n{3}$ when 
$M_1 =$ 175.7 and 170.9 GeV, respectively for $M_2 =$ 100 and 150 GeV. \\
Two additional remarks are in order. First, the vertical band of scenario 
(\ref{deg-sol-C}) is clearer and corresponds to an higher degree 
of degeneracy. This was explained above, in connection with the relation
of this scenario with the limit $\sthwq \to 0$. Second, both the 8 GeV 
and the 15 GeV contour have a bulge where they change direction, along the 
diagonal, towards lower values of $M_1$ and $M_2$. Even more visible is 
this effect if one observes the 25 GeV contour. This corresponds to the 
region around the point in the ($M_1, M_2$) plane where the further case 
of $m_1^o = m_2^o$ degeneracy we singled out is realized. 
This is given by Eq.(\ref{deg-sol-alfuno}), that, in the special case 
$\tgb = 1.2$, gives $M_1 \simeq M_2 \simeq -\mu \sindb = 179.4$. 
Some degeneracy is still present along the diagonal $M_1 = M_2$, to the left 
and below the highly degenerate region $M_1 \simeq M_2 \simeq -\mu$, although 
to a minor degree than in the horizontal and vertical bands. 
The degeneracy scenario (\ref{deg-sol-alfuno}) is more crucial in 
Fig.~\ref{N2-N1mass_mu-2Mz_tgb1p2}(b), where $\tgb$ is far away from 1 and 
the other degeneracy scenarios cannot be realized. Here the 8, 15 and 25 GeV 
contours surround the degeneracy point $M_1 = M_2 = -\mu \sindb = 85.8$ GeV 
and extend along the $M_1 = M_2$ diagonal in both the directions. 
Furthermore, note that this point does corresponds to a case of exact 
degeneracy, while the ``median'' lines of the horizontal and vertical bands 
in Fig.~\ref{N2-N1mass_mu-2Mz_tgb1p2}(a) do not, since $\tgb$ is only
approximately equal to 1. 
It is useful to check that the vertical line of approximate degeneracy 
[see, for instance, the one for $M_1 \simeq 175$-180 GeV in Fig.~5(a)] is
directly related to the mass-level crossing of two slightly mixed  
lightest neutralinos along the same line (see, for instance, the
behavior of A and B in Fig.~2). \\  
We will see in the next section that the degeneracy along the 
$M_1 = M_2$ diagonal will translate in explicit effects on BR(\nrad) 
for $\tgb$ well above 1, while, in the low $\tgb$ case, they will be 
mixed with and partly hidden by the dynamical enhancement. 
A final comment on Fig.~\ref{N2-N1mass_mu-2Mz_tgb1p2} is about the 
gaugino mass unification. We can see that is not possible in the unified 
case to realize a $\n{1}$--$\n{2}$ approximate degeneracy at a level 
of less than 25 GeV mass difference, unless $M_2 \gtap 300$ GeV and $\tgb$ 
is small (besides the very low $M_{1,2}$ region). 

\begin{figure}[h]
\centerline{ 
\epsfxsize=0.55\textwidth 
\epsffile{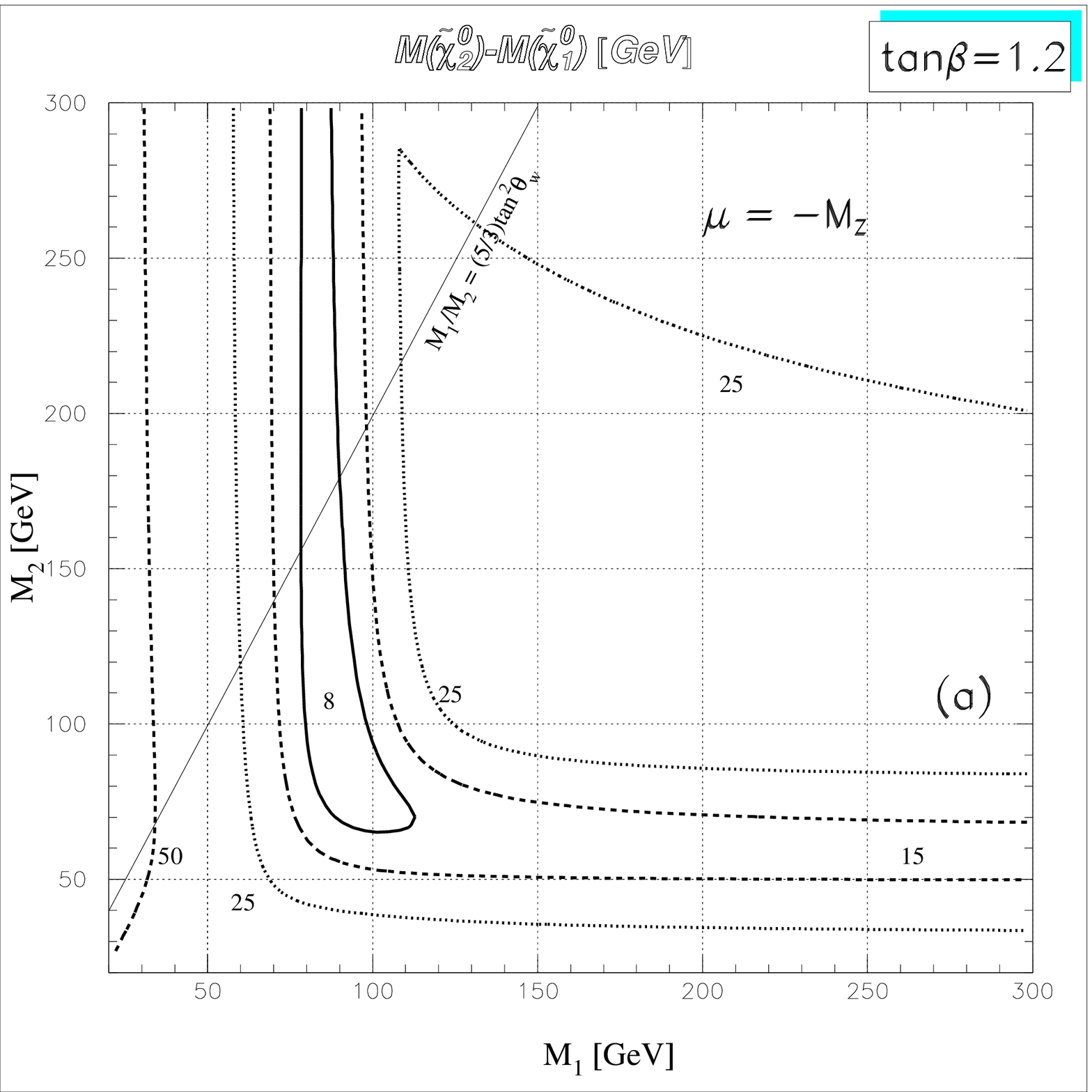}
\hspace{0.5cm} 
\epsfxsize=0.55\textwidth 
\epsffile{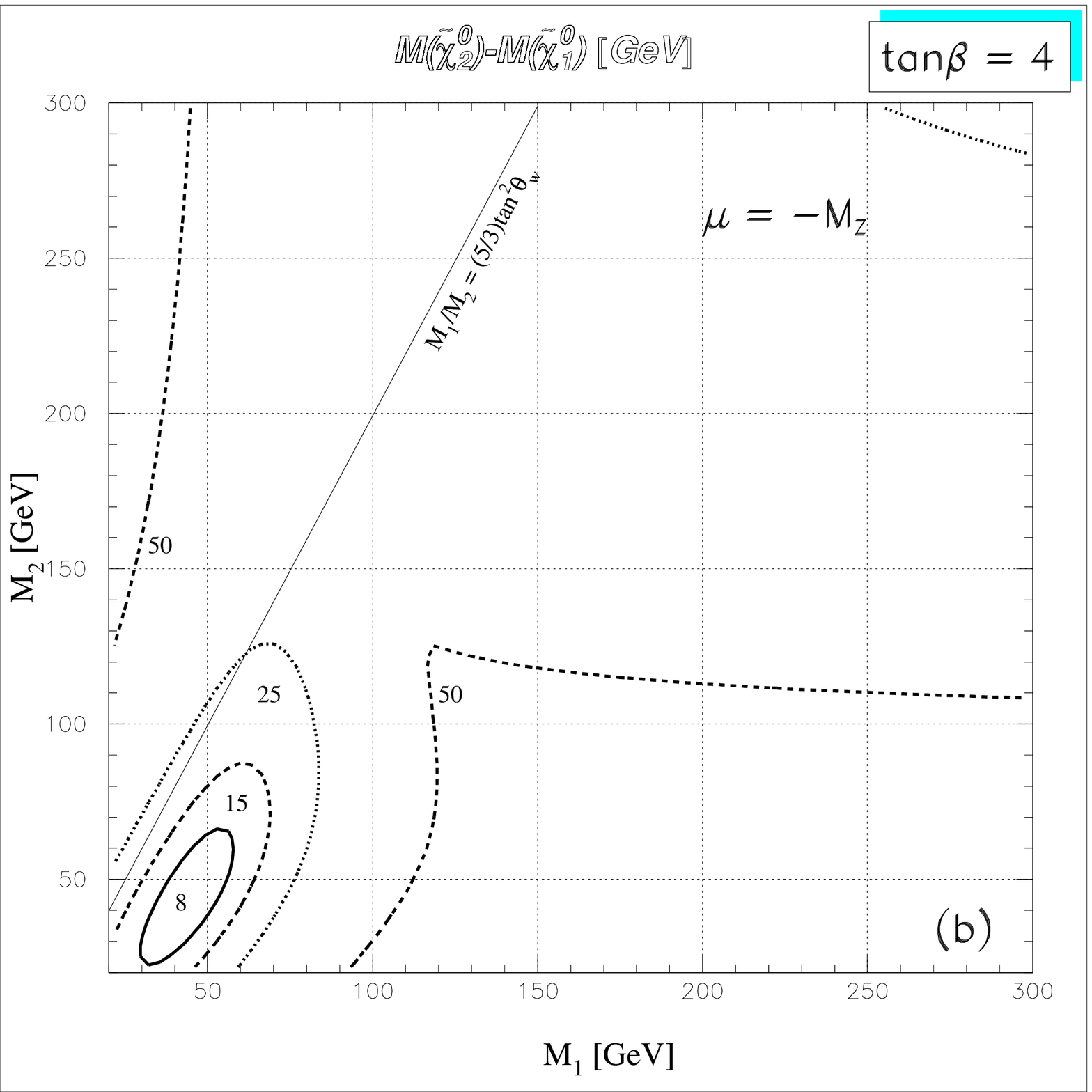}
}
\caption{The same as in Fig.~\protect\ref{N2-N1mass_mu-2Mz_tgb1p2}, but 
for $\mu = -M_Z$.}
\label{N2-N1mass_mu-Mz_tgb1p2}
\end{figure} 
\noindent 
Fig.~6 shows how the general picture evolves when $\mu$ goes from 
$-2M_Z$ to $-M_Z$. The behavior of ($\mn{2} - \mn{1}$) is qualitatively 
similar to the previous case, although the regions of strong degeneracy 
are shifted towards lower values of $M_{1,2}$. Also, it is now possible
to achieve interesting degeneracy scenarios for which the gaugino-mass 
unification holds and $M_{1,2}$ are in a range of interest for present
collider physics. 

Before coming to the numerical study of BR(\nrad), it is useful to note that 
the dynamical and kinematical mechanism can be present at the same time 
in special cases and work together to enhance BR(\nrad). For instance, 
a moderate degeneracy is needed between $\mn{1}$ and $\mn{2}$ for the
kinematical mechanism to be effective, when $\n{1}$ and $\n{2}$ also have 
a definite different composition and vice versa. 
Also note that the general necessary conditions 
(i.e. $\tgb \simeq 1$, $M_1 \simeq M_2$, with $\mu < 0$) are the 
same for both mechanisms, and this strengthens the enhancement effect.   

\section{Numerical analysis of BR(\nrad)}

On the basis of the results presented in Secs.~3 and 4 on the
enhancement regimes for the radiative $\n{2}$ decay, we are now ready
to explain the nontrivial behavior of the corresponding BR in the \susy\
parameter space. One of the main findings will be the existence of 
significant regions of this space, of interest for collider physics, 
where \nrad\ is the dominant $\n{2}$ decay.
Following the previous discussion, we present the BR(\nrad)
in the $(M_1,M_2)$ plane, for fixed values of $\tgb$ and $\mu$.
We also discuss the BR dependence on the scalar masses.

First of all, one has to set the parameter regions already excluded
by the experimental search. We recall that the usual analysis
implies the condition $M_1/M_2= (5/3) \tan^2\theta_w$.
Relaxing the latter, the definition of the exclusion
regions gets of course more involved. \\  
We considered the following bounds from LEP1 data on the $\Z$ line 
shape and on neutralino direct searches \cite{PDG,Feng,L3}:  
\beast 
\Gamma_{\rm TOT}(\Z \to {\rm SuSy}) & < & 23   \; \; {\rm MeV} \; , \\ 
\Gamma_{\rm inv}(\Z \to {\rm SuSy}) & < &  5.7 \; \; {\rm MeV} \; , \\ 
{\rm BR} (\Z \to \n{1} \n{2})       & < &  3.9 \times 10^{-6}  \; , \\ 
{\rm BR} (\Z \to \n{2} \n{2})       & < &  3.9 \times 10^{-6}  \; , 
\eeast  
where we took into account not only the $\n{1}\n{1}$ contribution to 
the $\Z$ invisible width, but all the channels $\Z\to\n{i}\n{j}$, \ 
$i,j = 1, \ldots,4$, \ with following invisible decays of the
produced heavier neutralinos 
$\n{i} \to \n{1} \nu \bar{\nu}(\ldots \nu \bar{\nu})$. \\
 
As for LEP1.5, we imposed the following limits from the direct searches of 
neutralinos/charginos during the runs at $\sqrt{s} = 130.3$ and 
136.3 GeV \cite{LEP130}
\beast 
\sigma_{\rm vis}\left[\epem \to 
\sum_{i,j}\left(\n{i}\n{j}, \; \cp{i}\cm{j} \right) \right]  
&  <  & 2 \; {\rm pb} \; \; {\rm at} \; \; \sqrt{s} = 130.3 \; {\rm GeV}
\; , \\ 
&     & {\rm if} \; 
(\mn{2} - \mn{1}) \; \; {\rm or} \; \;  
(\mc{1} - \mn{1}) > 10 \; {\rm GeV} \; , \\  
\sigma_{\rm vis}\left[\epem \to 
\sum_{i,j}\left(\n{i}\n{j}, \; \cp{i}\cm{j}\right) \right]  
&  <  & 2.4 \; {\rm pb} \; \; {\rm at} \; \; \sqrt{s} = 136.3 \; {\rm GeV} 
\; , \\ 
&     & {\rm if} \; 
(\mn{2} - \mn{1}) \; \; {\rm or} \; \; 
(\mc{1} - \mn{1}) > 10 \; {\rm GeV} \; , 
\eeast 
which corresponds to not allowing more than ten total visible events from
neutralino/chargino production at LEP1.5. Here too, the branching
fractions into visible final states of the produced particles have been 
taken into account while computing $\sigma_{\rm vis}$. 
The above LEP1.5 limits translate into an approximate bound
on the chargino mass $\mc{1} \gtap 65$ GeV, when the sneutrino is 
not too light ($m_{\tilde{\nu}} \gtap 200$ GeV) and there is enough phase 
space available to ensure the presence of rather energetic particles among the 
chargino decay products: $(\mc{1} - \mn{1}) > 10$ GeV. 

A few comments are in order. By applying the above experimental
constraints, one finds that for rather small values of $\tan\beta$
and $0\ltap\mu\ltap M_Z$  wide regions in the plane
$(M_1,M_2)$ are excluded. For instance, at $\tan\beta=1.2$
and $\mu=M_Z$, the area with $M_1,M_2\ltap 180$ GeV is forbidden.
The regions with positive $\mu$ are, however, not much relevant for the
radiative $\n{2}$ decay, as we know from Secs.~3 and 4. 
For low $\tgb$, negative $\mu$ and small $|\mu|$ (around 50 GeV)
one gets highly non-trivial exclusion regions (see below).
On the other hand, whenever the chargino search turns out to be the most 
effective means of constraining the plane $(M_1,M_2)$, the mass limit
and the forbidden region tend to get independent of $M_1$. The possible 
relevance of neutralino searches can translate, instead, in more
involved bounds, depending on $M_1$.

\begin{figure}[h]
\centerline{ 
\epsfxsize=0.55\textwidth 
\epsffile{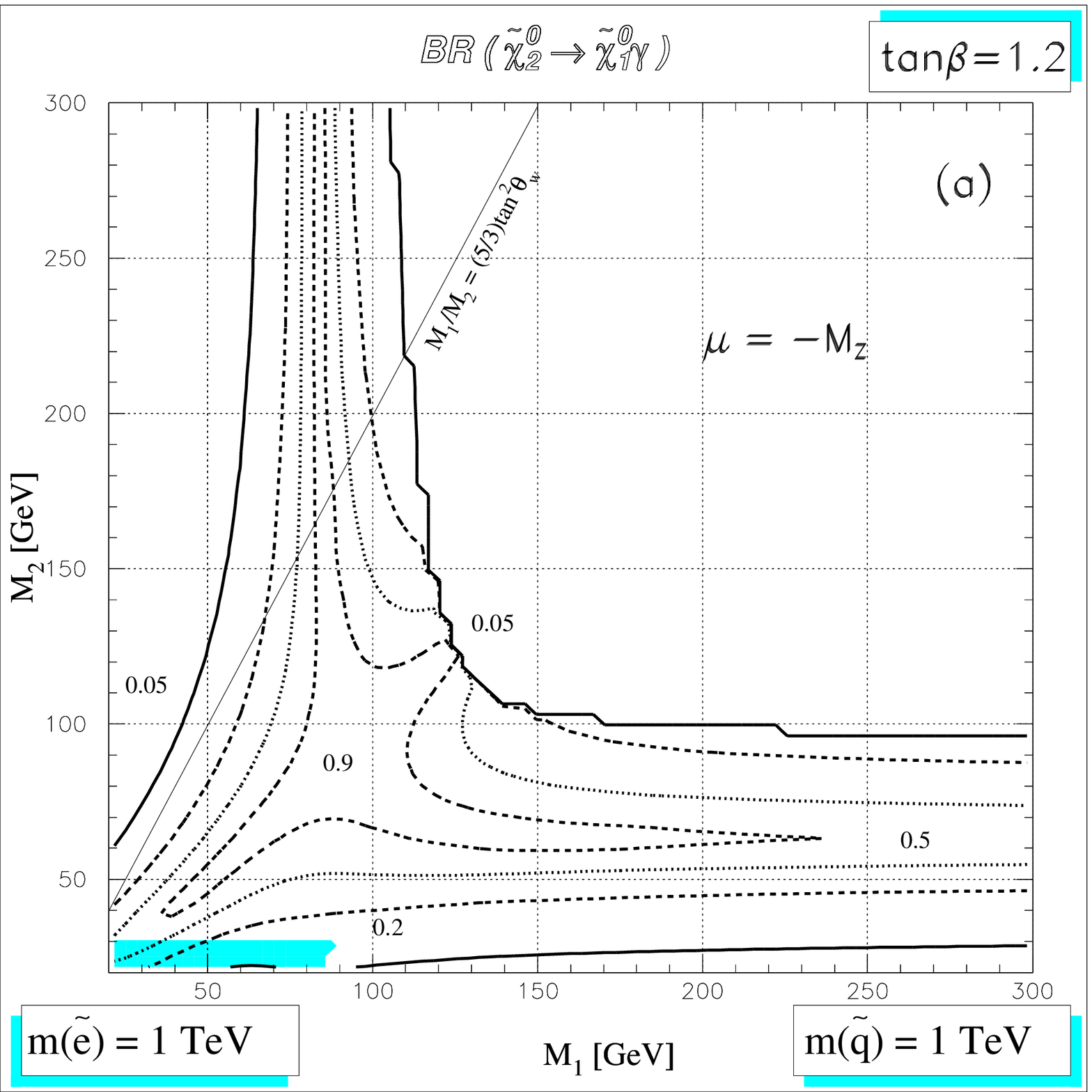}
\hspace{0.5cm} 
\epsfxsize=0.55\textwidth 
\epsffile{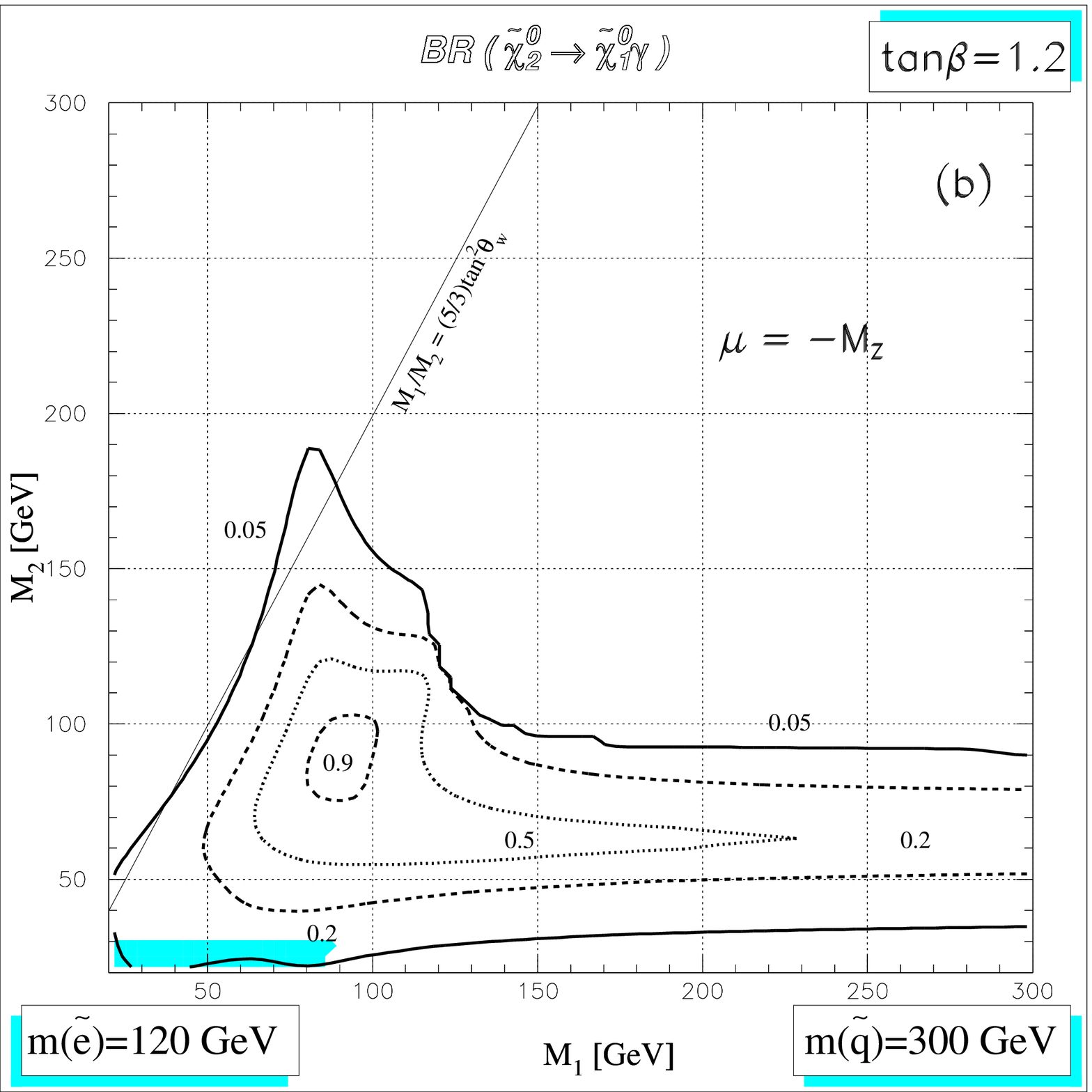}
}
\caption{Contour plot for the branching ratio of the neutralino 
radiative decay. In (a), all the sfermions are taken degenerate
with mass equal to 1 TeV. In (b), the left and right charged slepton 
masses are taken degenerate and equal to 120 GeV; the sneutrino mass 
is calculated by using the SU(2) sum rule; the squark masses are taken 
all degenerate and equal to 300 GeV. 
The Higgs sector masses and couplings are set by $m_A = 300$ GeV. 
The shaded region is excluded by LEP1 and LEP1.5 data.}  
\label{BRad_mu-91_tgb1p2}
\end{figure} 

In Fig.~7, we show the BR(\nrad) for $\tgb = 1.2$ and
$\mu = -M_Z$, i.e., in a regime where both the dynamical and kinematical
enhancement can be realized. The shaded region is excluded by LEP1-1.5 data. 
In this figure and in all the following ones, we calculate the Higgs
sector masses and couplings by assuming $m_A = 300$ GeV. 
In Fig.~7(a), the kinematical enhancement is optimized by the large value 
(1 TeV) assumed for the sfermion masses. We find a significant area
in the plane $(M_1,M_2)$ where BR(\nrad)$ > 90\%$. Its shape
can be straightforwardly explained by putting together
the information from Fig.~3, on the physical {\it purity} of the two 
lightest neutralinos, and Fig.~6(a), on the neutralino mass degeneracy. 
It is interesting to note in Fig.~7(a) how much, along the 
$M_1 \simeq M_2$ diagonal, different enhancement mechanisms can be
effective and can contribute to
a large BR(\nrad)$ \gtap 90\%$, depending on the $M_{1,2}$ values.
For instance, when $M_1 \simeq M_2 \simeq 40$ GeV, the two lightest
neutralinos have widely different masses [cf. Fig~6(a)] and the 
kinematical enhancement is not operative. Furthermore, since $\tgb$ is not 
exactly 1, $\n{2}$ has comparable $\Zino$ and $\hino{b}$ components, 
spoiling the full dynamical enhancement. However, since the sfermions are 
very heavy, it is sufficient the presence of a nearly pure 
$\n{1} \simeq \phino$ to largely deplete the only tree-level 
decay channel left (through $\Z$ exchange) and to give rise to very large 
BR(\nrad) values ({\it reduced} dynamical suppression, cf. Sec.~2).
Of course, after lowering the sfermion masses, 
this enhancement regime does not survive [cf. Fig.~7(b)]. 
Proceeding towards higher $M_{1,2} \simeq -\mu\sindb$ values on the
diagonal, one has two enhancement mechanisms getting effective at the
same time. First, the 
$\hino{b}$ component of the $\n{2}$ grows [cf. Fig~3(a)], giving rise to 
a ``full'' dynamical enhancement. Second, $\mn{1}$ and $\mn{2}$ get
closer and closer [cf. Fig.~6(a)] and the kinematical mechanism becomes 
effective as well. The most favorable situation is then realized in this
region. For larger values of $M_{1,2}$ in the [100, 120] GeV range,  
the dynamical enhancement dominates, but with $\n{1} \simeq \hino{b}$ 
and $\n{2} \simeq \phino$ [cf. Fig.~3(b)]. Finally, when $M_1$ and $M_2$ 
are both $\gtap 120$ GeV, large BR(\nrad) values cannot be achieved 
anymore. 

\begin{figure}[t]
\centerline{ 
\epsfxsize=0.55\textwidth 
\epsffile{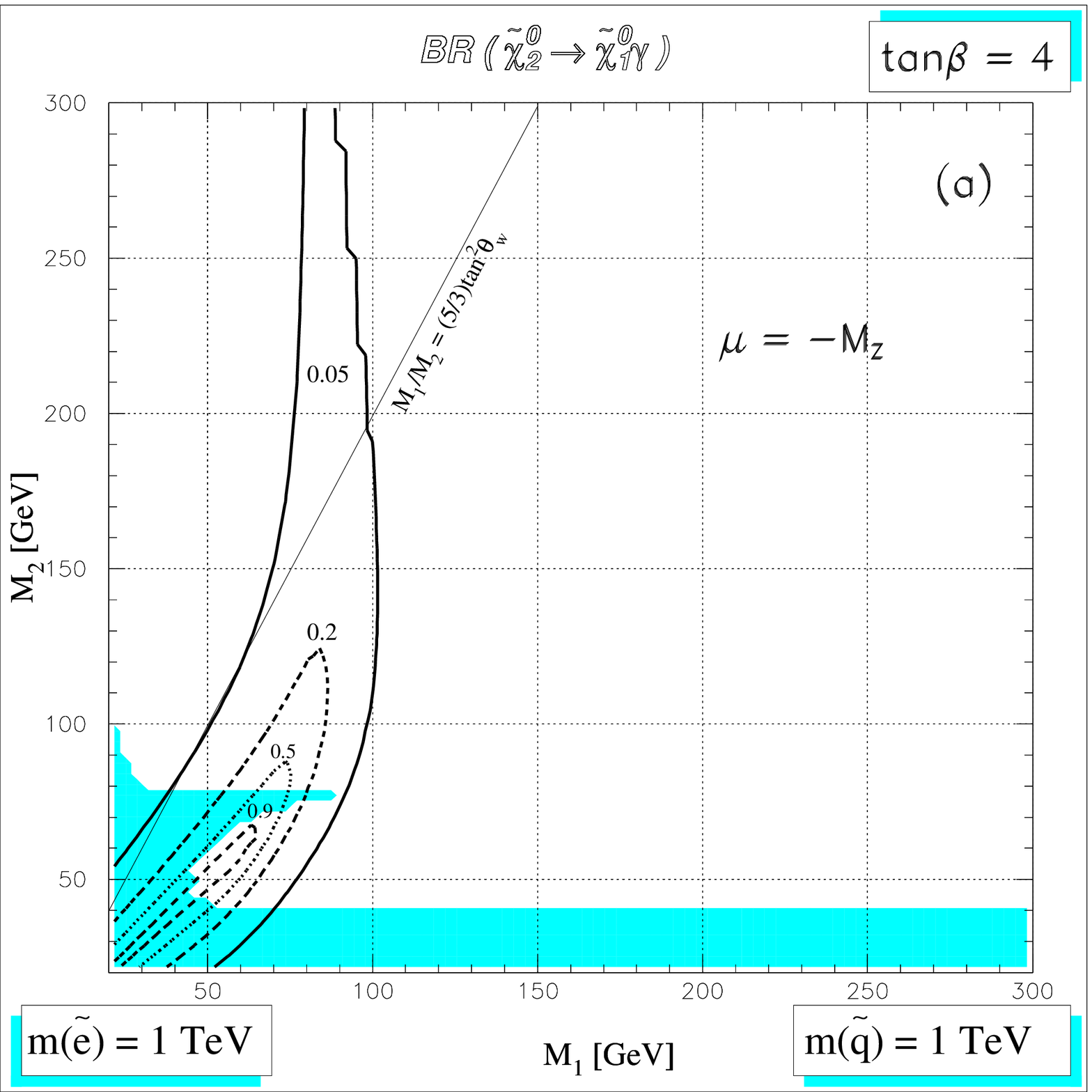}
\hspace{0.5cm} 
\epsfxsize=0.55\textwidth 
\epsffile{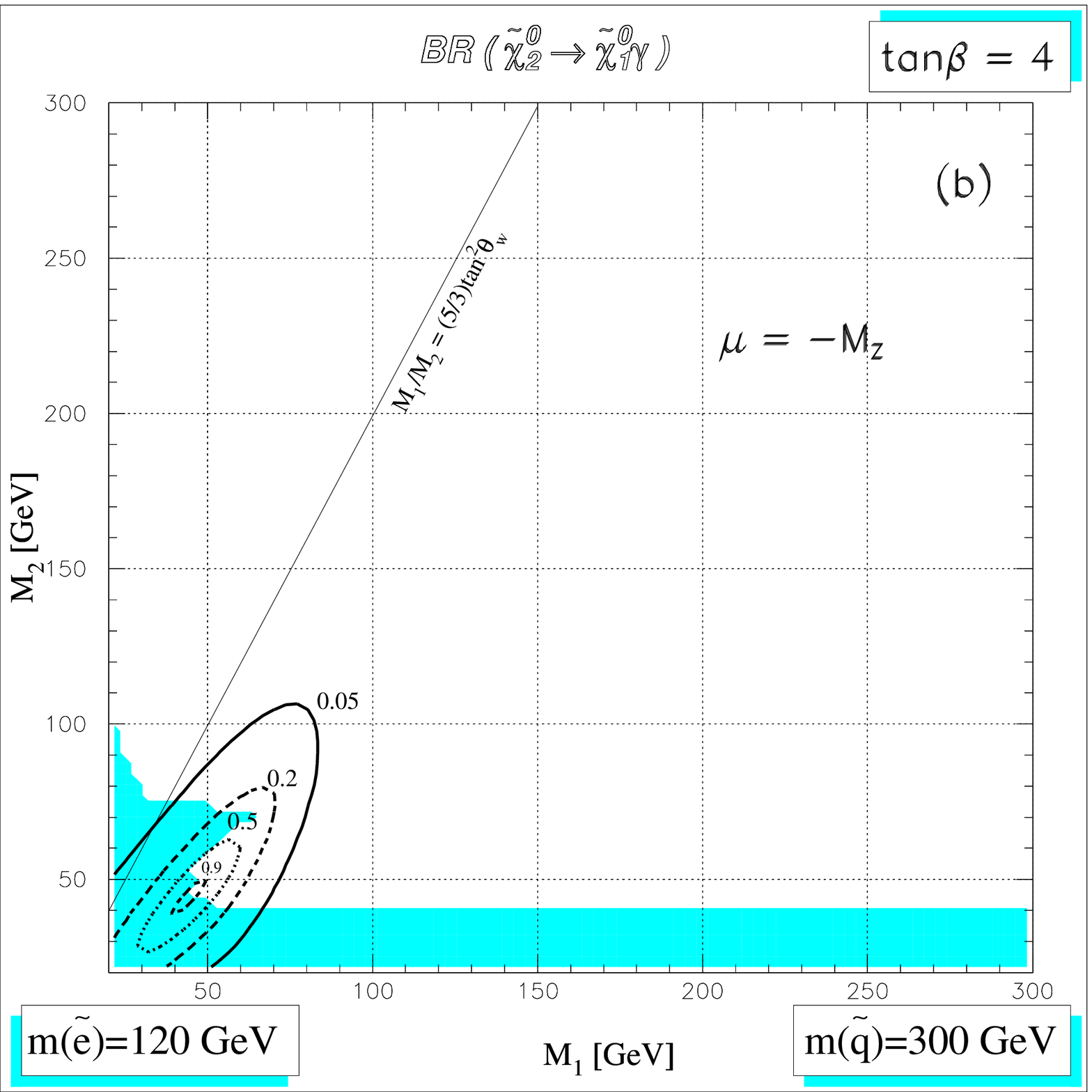}
}
\caption{The same as in Fig.~\protect\ref{BRad_mu-91_tgb1p2}, 
but for $\tgb = 4$.} 
\label{BRad_mu-91_tgb4}
\end{figure} 

The relative importance of the various kinematical enhancement scenarios
can be guessed by comparing Fig.~7(a) with Fig.~7(b).
In the latter, lighter scalars, that tend to reduce the effect of the
kinematical suppression in the $\n{2}$ tree level decays,
are assumed. Nevertheless, a strong radiative enhancement
mainly due to a dynamical suppression of the tree-level decays
is still present for $M_1 \simeq M_2 \simeq -\mu$.
Note that the kinematical enhancement which is expected from the scenario 
(\ref{deg-sol-AB})$_+$ [cf. the horizontal band of approximate degeneracy 
in Fig.~6(a)] is only slightly influenced by the change of the sfermion mass 
values, while the scenario (\ref{deg-sol-C}) (vertical band) corresponds
to higher values of BR(\nrad) for heavy sfermion masses, and is much
more sensitive when changing these parameters.  
This happens in spite of the greater degree of neutralino mass
degeneracy corresponding to the scenario (\ref{deg-sol-C}).
The reason is that in the case (\ref{deg-sol-AB}), in addition to the 
kinematical suppression of the tree-level decay widths, one has a rather
large value of the radiative decay width, due to the presence of light
charginos (low $M_2$) in the $\Wpm/\cmp{}$ loops, irrespective of the sfermion
masses. For instance, the point $M_1 = 180$ GeV, $M_2 = 60$ GeV in Fig.~7(a)
corresponds to a total $\n{2}$ decay width of about 100 eV and only about 
5\% of it is due to tree-level decays (in particular, cascade decays with 
$\mn{2} = 97$ GeV, $\mc{1} = 93$ GeV, and $\mn{1} = 85$ GeV). In
contrast, in the ``unified'' point $M_1 = 85.2$ GeV, $M_2 = 170$ GeV,  
one has $\Gamma_{\rm TOT}(\n{2}) \simeq \Gamma(\n{2}\to\n{1}\gamma) \sim 1$ eV 
for 1 TeV sfermion masses, while the total $\n{2}$ width approaches 50 eV 
for the lighter sfermion masses used in Fig.~7(b), about 90\% of it
coming from tree-level light-slepton exchange channels. 

\begin{figure}[t]
\centerline{ 
\epsfxsize=0.55\textwidth 
\epsffile{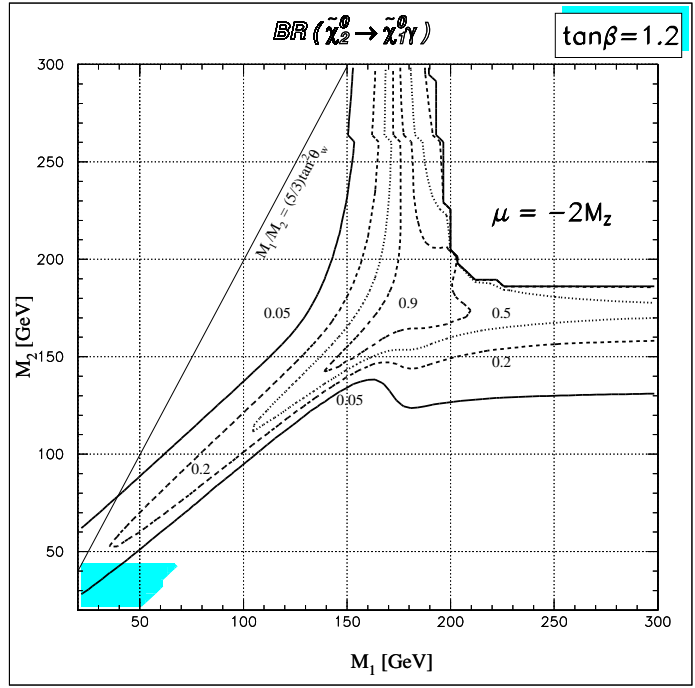}
} 
\caption{Contour plot for the neutralino radiative decay BR in the 
case $\tgb = 1.2$ and $\mu = -2M_Z$ . 
The sfermion masses are all taken degenerate and equal to 1 TeV and 
$m_{A} = 300$ GeV. The shaded region is excluded by LEP1-1.5 data.}
\label{BRad_mu-273_tgb1p2}
\end{figure}

A final remark on Fig.~7 is that there are regions where neither the
kinematical nor the dynamical BR(\nrad) enhancement are fully effective,
since the regimes for the \susy\ parameters we outlined in Secs.~3 and 4 
are only realized with a large approximation. Nevertheless, the combined 
effect of the two mechanisms can still give rise to large values of BR(\nrad). 

When $\tgb$ rises (for instance, $\tgb=4$ 
is assumed in Fig.~8) the physical {\it purity} of neutralinos decreases
(see the corresponding Fig.~4), and mainly
effects connected to the kinematical enhancement in the scenario 
(\ref{deg-sol-alfuno}) survive. This can be easily verified by comparing
Fig.~8 with Fig.~6(b). The value of the sfermion masses is then a relevant
parameter. Assuming lighter scalars [Fig.~8(b)] considerably reduces
the region of large radiative BR with respect to the regime where
the sfermion exchange in the tree-level $\n{2}$ decay is suppressed
by heavy scalars [Fig.~8(a)].

\begin{figure}[t]
\centerline{ 
\epsfxsize=0.55\textwidth 
\epsffile{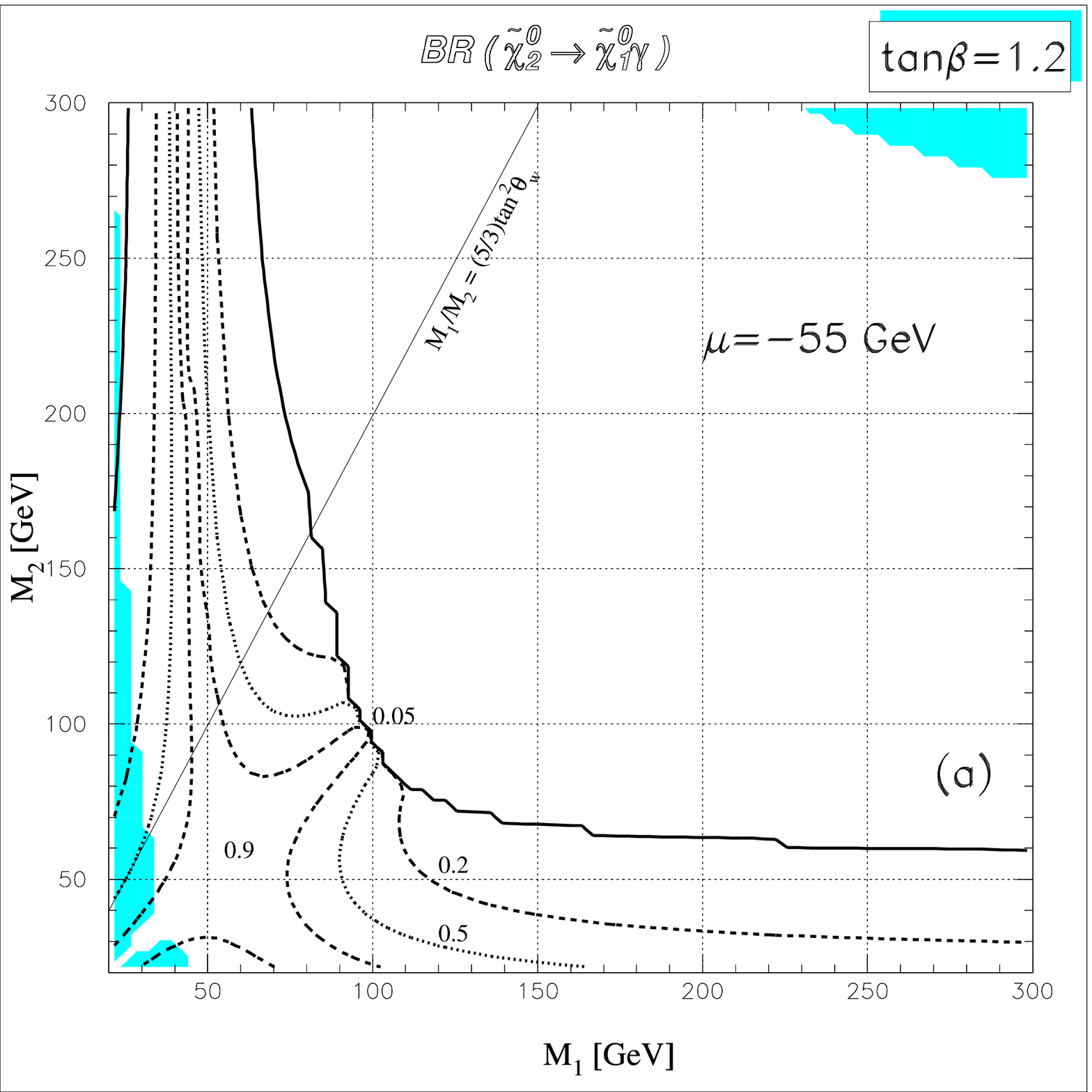}
\hspace{0.5cm} 
\epsfxsize=0.55\textwidth 
\epsffile{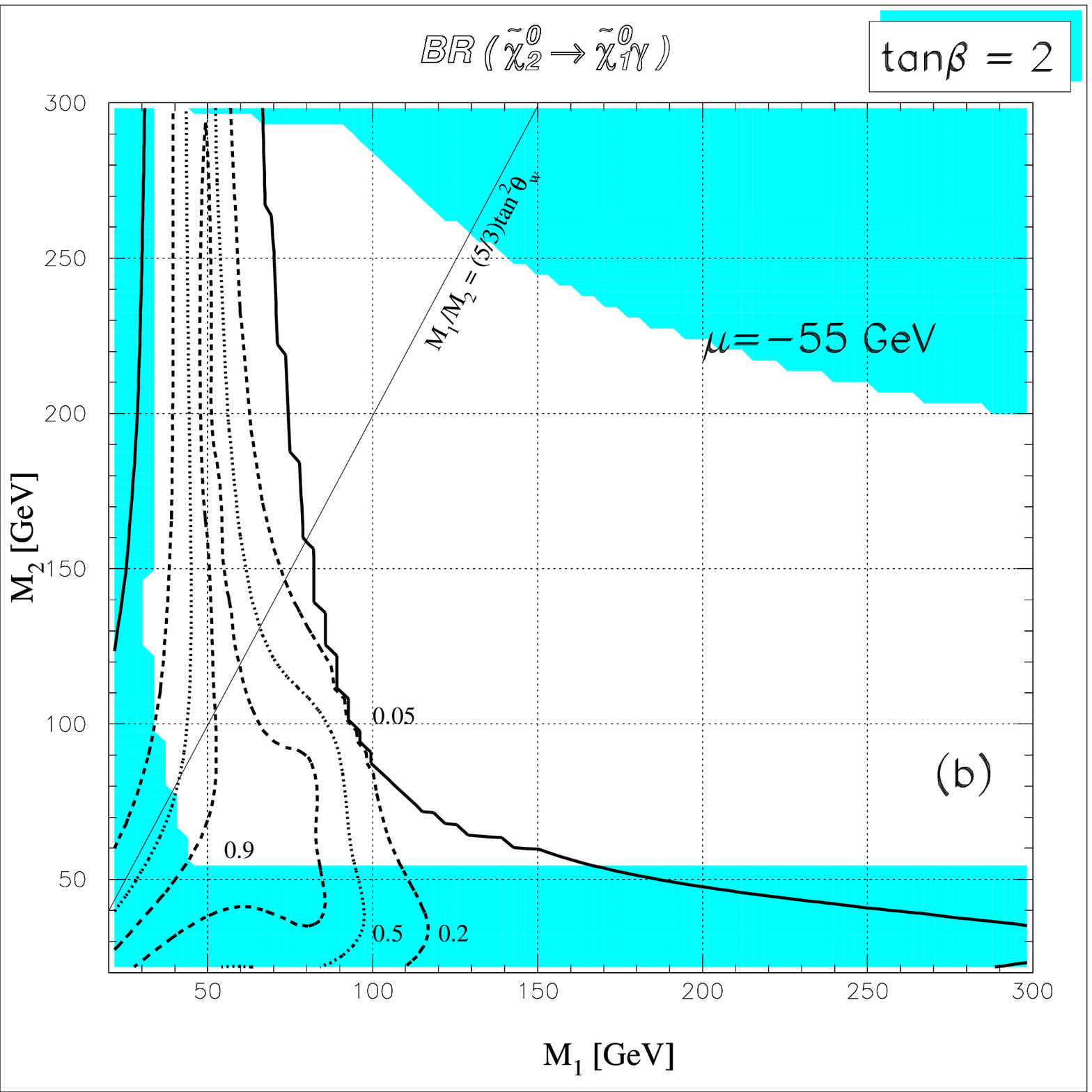}
}
\caption{The same as in Fig.~9, but for $\mu=-55$ GeV  
and $\tgb = 1.2$ (a) and 2 (b).}
\label{BRad_mu-55}
\end{figure}

We consider now the effect of varying the parameter $\mu$.
From now on, we will keep the sfermion masses rather heavy
[i.e., we set $m_{\tilde{e}} = m_{\tilde{q}} = 1$ TeV] in order 
to optimize the BR(\nrad) enhancement.
In Fig.~9, the effect of decreasing $\mu$ down to $-2M_Z$
is shown for $\tgb = 1.2$. 
\begin{figure}[h]
\centerline{ 
\epsfxsize=0.55\textwidth 
\epsffile{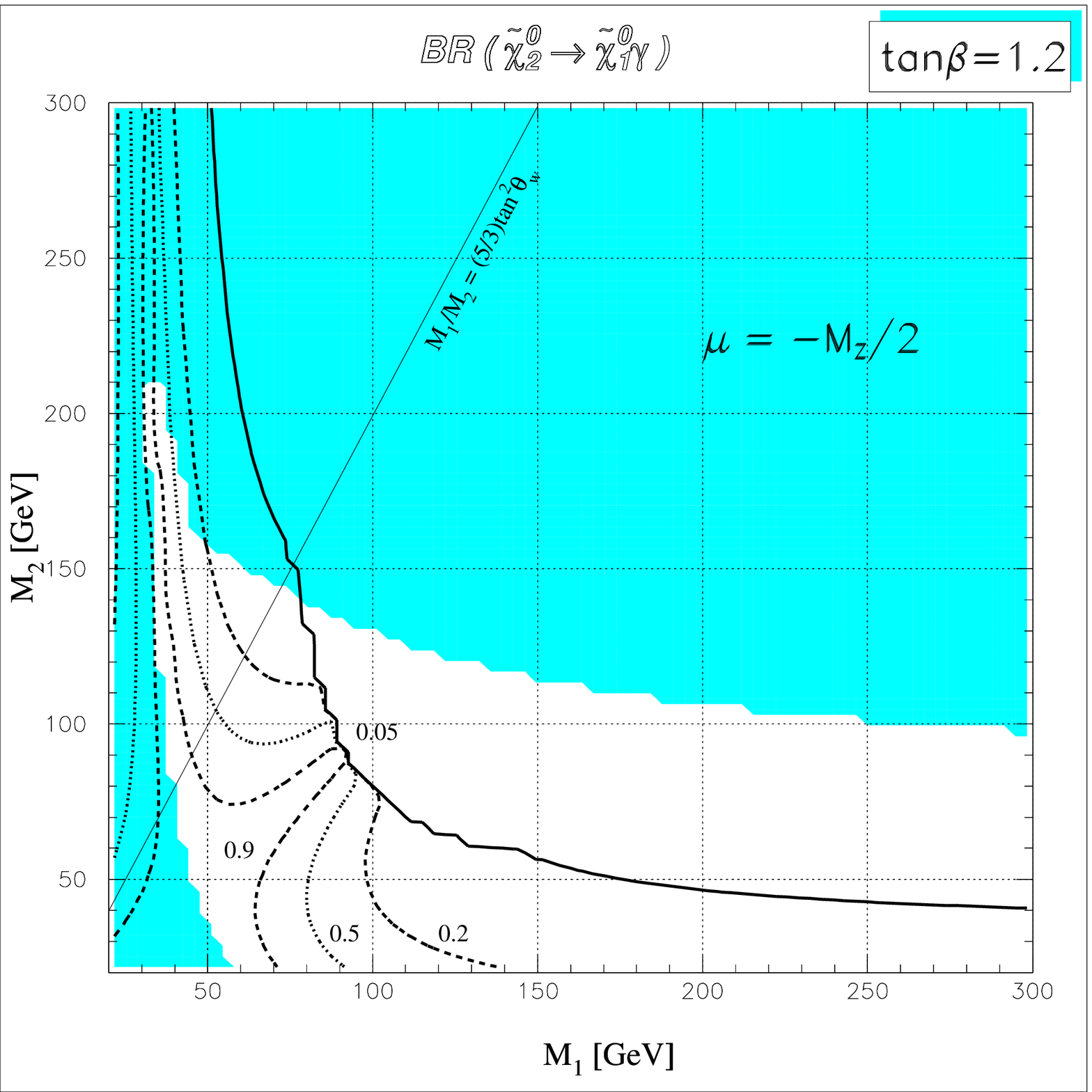} 
} 
\caption{The same as in Fig.~9, but for $\mu = -M_Z/2$
and $\tgb=1.2$.}
\label{BRad_mu-45_tgb1p2}
\end{figure}
Again, some insight of the BR behavior can be gained by 
looking back at Fig.~2 (physical ``purity" of the neutralinos)
and Fig.~5(a) (neutralino mass degeneracy).
The area around $M_1 \simeq M_2 \simeq -\mu$ is again particularly
promising. However, quite large radiative BR's can be also achieved in three
strips of the ($M_1,M_2$) plane, corresponding to different 
kinematical enhancement scenarios [cf. Fig.~5(a)]. Some effects from 
the {\it reduced} dynamical mechanism with heavy sfermion masses are also 
present. 

Keeping $\mu$ in the negative range, we now go to larger $\mu$ values, 
which are of particular interest for explaining the \eegg\ event at the
TeVatron. In Figs.~10(a) and 10(b), the results for $\mu=-55$ GeV at 
$\tgb = 1.2$ and 2, respectively, are shown. 
While a large BR(\nrad) can still be obtained
comfortably, we can observe that for increasing $\tan\beta$ the regions 
excluded by present data extend further and further.

Going to $\mu=-M_Z/2$ at $\tan\beta=1.2$ 
(cf. Fig.~11) has the effect of a moderate
shifting of the large BR regions down to smaller $M_1,M_2$ with respect to 
Fig.~10(a), accompanied again by a drastic reduction of the parameter space 
allowed by the experimental data. 

\section{Top-squark sector influence on BR(\nrad)} 

As long as $\mc{1} \ltap M_W$, the main contributions to the
$\n{2}$ radiative decay width come from the $\Wpm/\cmp{}$ loops
(cf. Fig.~1), unless one considers scenarios with light top-squarks, 
i.e. 60 GeV $\ltap m_{\tilde{t}} \ltap m_t$ (throughout this paper we assume 
$m_t = 175$ GeV). In the latter case, the amplitudes corresponding to 
$t/\tilde{t}$ loops are non-negligible and can interfere destructively 
with the $\Wpm/\cmp{}$ loops, hence decreasing the radiative decay width. 
If one assumes the mass of the heavy top-squark $\tilde{t}_2$ 
sufficiently larger than $m_t$, then the bulk of this effect comes from 
the light top-squark $\tilde{t}_1$. 
Note that the particular values of the other sfermion masses have an 
important influence just on the tree-level neutralino decay widths, hence 
affecting only the BR of the radiative decay. The opposite happens for the 
top-squark parameters, which directly influence the radiative width, and
are not involved in the tree-level decays in the parameter ranges
considered here. 
In the previous sections, we assumed the two top-squarks  
$\tilde{t}_1$ and $\tilde{t}_2$ degenerate, with a mass equal to 
all the other squark masses. Here, we relax this simplification and 
consider the effects of assuming different values for $m_{\tilde{t}_{1}}$
and $m_{\tilde{t}_{2}}$ and, in particular, the possibility of a 
rather light $\tilde{t}_1$.  
In addition, since the mass eigenstates $\tilde{t}_{1,2}$ can be  
superpositions of the interaction eigenstates $\tilde{t}_{L,R}$, we 
also take into account the effect of varying the top-squark mixing angle
$\theta_t$, defined by:
$\tilde{t}_1 =  \cos\theta_t \tilde{t}_L + \sin\theta_t \tilde{t}_R$; \ \ 
$\tilde{t}_2 = -\sin\theta_t \tilde{t}_L + \cos\theta_t \tilde{t}_R$. 

\begin{figure}[t] 
\centerline{ 
\epsfxsize=0.55\textwidth 
\epsffile{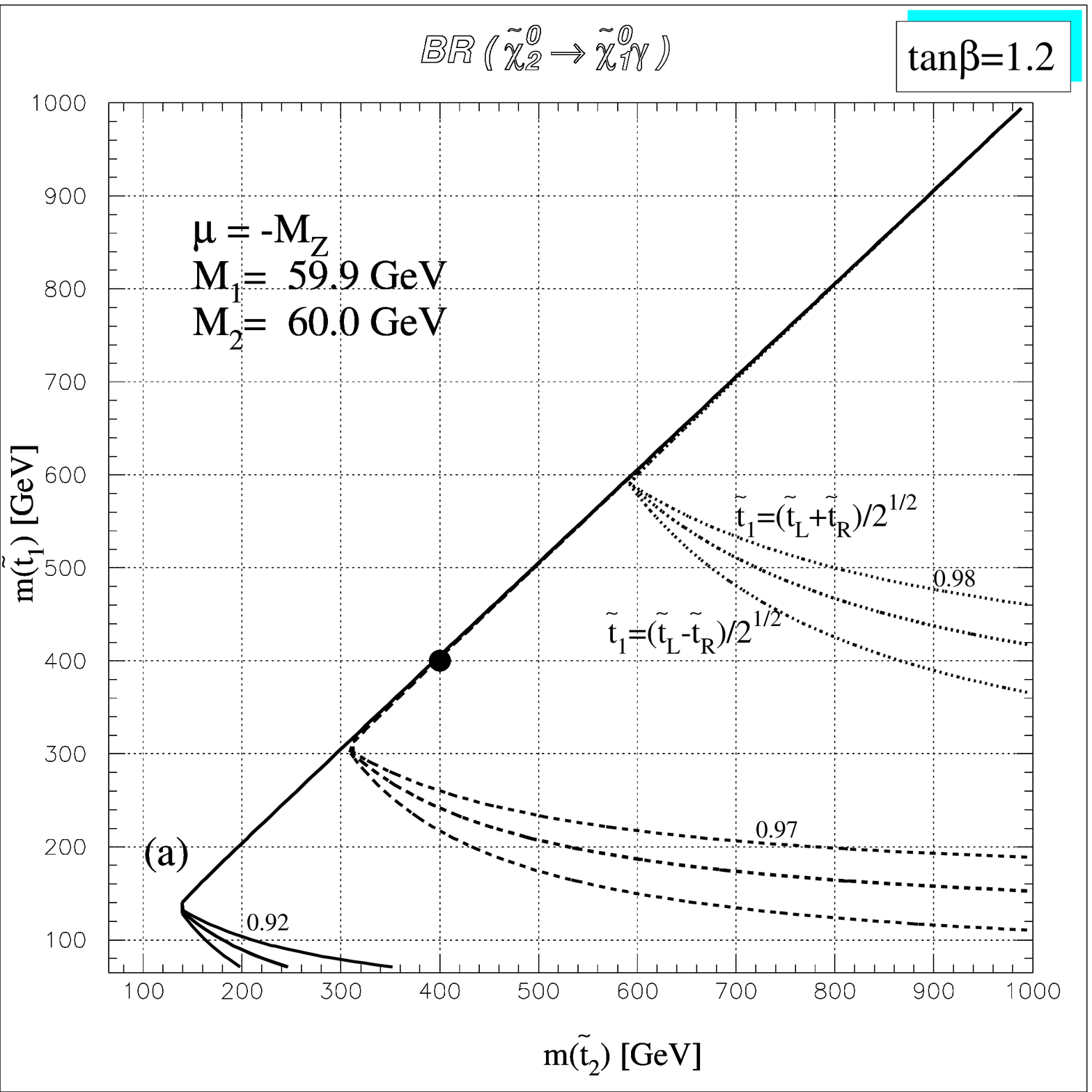}
\hspace{0.5cm} 
\epsfxsize=0.55\textwidth 
\epsffile{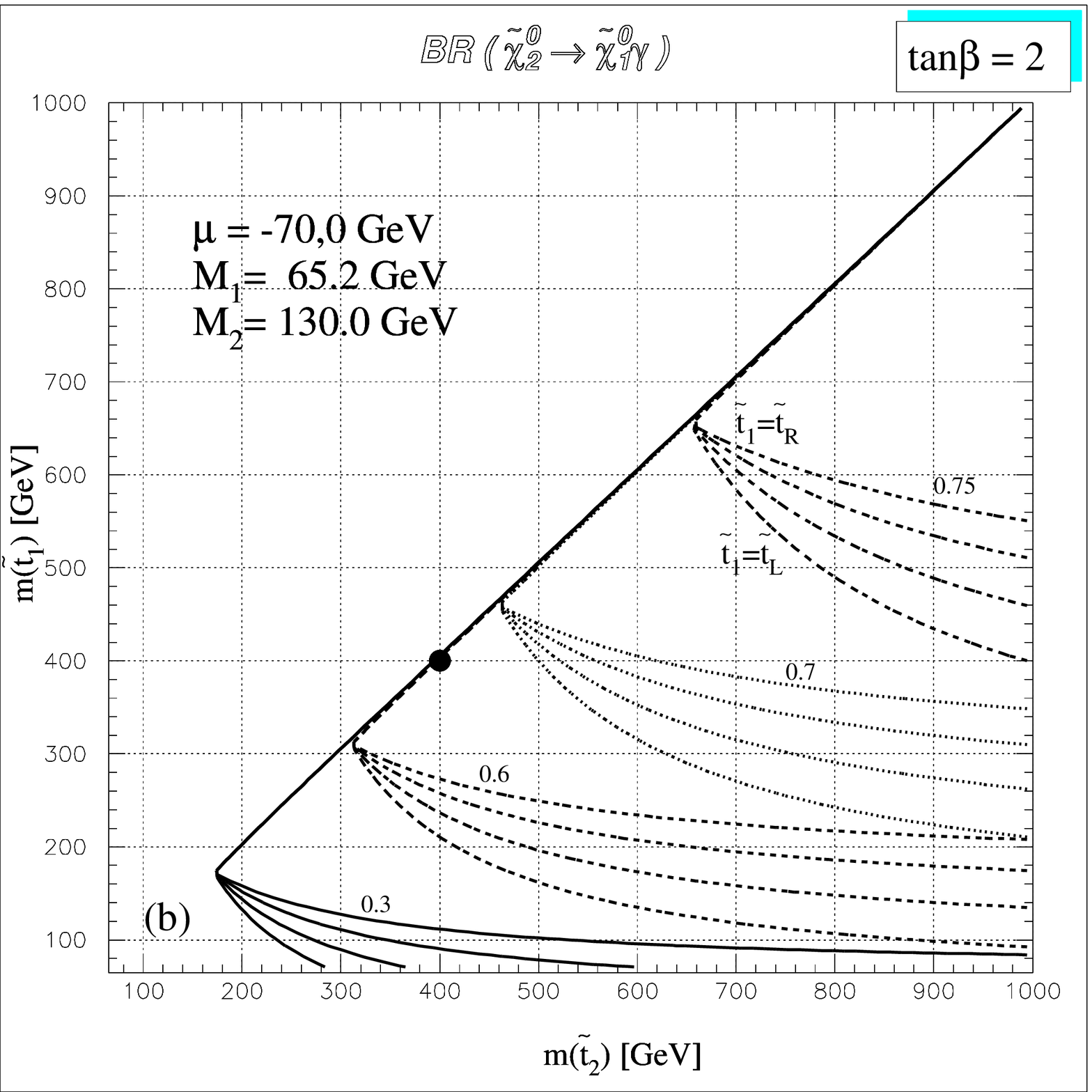}
}
\caption{Contour plot for BR(\nrad) in the 
($m_{\tilde{t}_1}, m_{\tilde{t}_2}$) plane. 
For each fixed value of BR(\nrad), four curves corresponding 
to different choices of the mixing angle $\theta_t$ are shown with lines
of the same style. 
Going from the lower curve towards the upper one, one has: \protect\\ 
(a) Typical dynamical enhancement scenario: \protect\\ 
$\theta_t = -\pi/4$, 0 (or $\pi/2$), $+\pi/4$;
\protect\\ 
(b) Special kinematical enhancement scenario with gaugino mass unification:
\protect\\  
$\theta_t = 0$, $-\pi/4$, $+\pi/4$, $+\pi/2$.  \protect\\ 
The other squark masses are taken degenerate and equal to 400 GeV and
the slepton masses are set at 300 GeV. Also, $m_{A^0} = 300$ GeV.
The big black dot corresponds to assuming the top-squark masses equal to the 
other squark masses, as in the previous sections.} 
\label{BRadvstop}
\end{figure}

In Fig.~12, we show the contour plot of BR(\nrad) as a function of the
two top-squark masses, for four values of the mixing angle $\theta_t$ and for 
two interesting choices of the neutralino sector parameters which
correspond to large-BR(\nrad)
scenarios with different characteristics. Here, we assume degenerate
slepton masses of 300 GeV and the other squark masses all equal to 400
GeV. The pseudoscalar Higgs mass is fixed at 300 GeV.  

Here we did not put any restriction on the $(m_{\tilde{t}_1}, 
m_{\tilde{t}_2}$) plane (besides a rough direct limit of about 60 GeV 
on the top-squark masses from LEP data) and we considered the physical masses 
$m_{\tilde{t}_i}$ as independent parameters. 
However, in the framework of the MSSM, even without precise unification 
assumptions in the scalar sector, one usually derives the physical
sfermion masses and the L-R mixing angles from the soft \susy\ breaking 
parameters (i.e.: $\tilde{m}_{q_L}$, $\tilde{m}_{u_R}$,
$\tilde{m}_{d_R}$, \ldots, $A_t$, $A_b$, \ldots) as well as from 
$\mu$ and $\tgb$ (cf., e.g., Ref.\cite{Barger}). 
Once the values of the squark masses of the
first two families are fixed to be roughly degenerate, e.g. at 400 GeV as
in Fig.~12, and assuming the mass parameters in the sbottom sector 
to have similar values as well
(in particular $\tilde{m}_{b_L} = \tilde{m}_{t_L}$), 
then it seems unnatural to build a
coherent model with the heavy top-squark $\tilde{t}_2$ considerably lighter 
than 400 GeV, especially when $\tilde{t}_1$ is very light. Similarly, 
the light top-squark is usually lighter than the other squarks. 
Furthermore, one has to take into account the influence of the top-squark  
sector on the light Higgs mass $m_h$ through the radiative corrections
(cf., e.g., Ref.\cite{Barger}). Especially for low $\tgb$   
and for moderate $m_A$, the presence of 
a very light $\tilde{t}_1$ ($m_{\tilde{t}_1} \ltap 70$-80 GeV) may
induce $m_h$ to fall below the present experimental limits (i.e. 44 GeV 
or roughly $\sin^2(\beta-\alpha)\times 60$ GeV, from Ref.\cite{PDG}), 
unless $\tilde{t}_2$ is very heavy ($m_{\tilde{t}_2} \gtap 1$ TeV). 
The value $m_A = 300$ GeV we used in this section generally makes the
latter problem negligible, but the discussion above still suggests 
that some regions of the top-squark mass plane we show in Fig.~12 may not 
correspond to a physically acceptable scenario. Nevertheless, having
this in mind, Fig.~12 and the following one provide useful hints to
evaluate the pattern of the top-squark sector influence on BR(\nrad). 

In Fig.~12(a), we present a typical case of dynamical enhancement 
($M_1 \simeq M_2 = 60$ GeV, $\mu = -M_Z$, and $\tgb = 1.2$), while, 
in Fig.~12(b), a special case of kinematical enhancement is considered. 
The latter (with $M_1 = 65.2$ GeV, $M_2 = 130$ GeV, $\mu = -70$ GeV, 
and $\tgb = 2$) does not optimize the kinematical mechanism, but
has two interesting features. First, the case (b) satisfies the gaugino
mass unification relation $M_1=\frac{5}{3}\tgwq M_2$. Second, the
value of $\tgb$ quite different from 1 gives rise to a sizeable mass 
splitting between the neutralinos ($\simeq 17$ GeV) and to rather
energetic photons, even after the radiative decay of a {\it soft} $\n{2}$. 
Furthermore, in the case (b) a certain amount of dynamical enhancement
is also present. 

For each fixed value of the radiative BR, we show how the contours move
when varying the mixing angle $\theta_t$. In particular, for a given  
BR(\nrad), we present four contours corresponding to 
$\theta_t = 0$, $-\pi/4$, $\pi/4$, and $\pi/2$. This ordering corresponds 
to going from the lower curve to the upper one with the same line style 
in Fig.~12(b), while, in Fig.~12(a), the lower curve is obtained for $\theta_t
= -\pi/4$, the upper one for $\theta_t = +\pi/4$, and the curve in the middle
corresponds to the almost degenerate contours for $\theta_t = 0$ and $\pi/2$. 
As a reference, we also show by a big black dot the scenario of complete
squark mass degeneracy (in particular $m_{\tilde{t}_{1,2}} = m_{\tilde{q}}
= 400$ GeV), assumed in the previous analysis. Under that hypothesis, one would
obtain BR(\nrad) $\simeq 97.5\%$ in the case (a) and BR(\nrad) 
$\simeq 65\%$ in the case (b). 
As anticipated, the radiative BR increases with the top-squark masses, showing 
a larger sensitivity to $m_{\tilde{t}_1}$. Indeed, when the mass 
splitting ($m_{\tilde{t}_2} - m_{\tilde{t}_1}$) is sizeable, BR(\nrad)
gets independent of the heaviest mass $m_{\tilde{t}_2}$. 

As for the physical composition of the top-squark mass eigenstates entering 
the radiative decay loops, Fig.~12(b) shows that a lightest top-squark
corresponding to a pure $\tilde{t}_R$ ($\theta_t = \pi/2$) is more effective
in reducing the radiative BR in the considered case of kinematical 
enhancement. Indeed, for a fixed $m_{\tilde{t}_1}$, 
taking $\tilde{t}_1 = \tilde{t}_R$ gives rise to a larger cancellation 
between the $t/\tilde{t}_1$ and the $\Wpm/\cmp{}$ loops than in the 
$\tilde{t}_1 = \tilde{t}_L$ case. The mixed cases corresponding to 
$\theta_t = \pm \pi/4$ fall in the middle of the two pure cases.  
In the dynamical enhancement scenario of Fig.~12(a), the behavior is somehow 
opposite. It turns out that taking a pure $\tilde{t}_L$ or $\tilde{t}_R$ 
state as the lightest top-squark can not be distinguished in BR(\nrad), 
while if $\tilde{t}_1$ is the symmetric (antisymmetric) combination 
$[\tilde{t}_L +(-) \tilde{t}_R]/\sqrt{2}$, the contributions from 
$t/\tilde{t}$ loops are maximized (minimized) and the same happens to
the larger destructive interferences with the $\Wpm/\cmp{}$ loops. 
This is because in the case (a) the two neutralinos are an almost 
pure photino and an almost pure $\hino{b}$, which both couple with the same 
strength to left- or right-type standard fermions and sfermions. 
As a consequence, the contributions from pure $t/\tilde{t}_L$ and 
$t/\tilde{t}_R$ loops are almost equal, while, if the mass eigenstate 
considered is an antisymmetric combination, its contribution to the
matrix element is negligible. When this is the case for $\tilde{t}_1$,
and $\tilde{t}_2$ is heavy, the whole top-squark sector does not 
contribute and the process is dominated by $\Wpm/\cmp{}$ loops. 
In the case (b), instead, the $\n{1,2}$ composition is more involved,
with sizeable $\Zino$ components, which distinguish between 
$f_L\tilde{f}_L$ and $f_R\tilde{f}_R$.  

Comparing Fig.~12(a) and 12(b), one notes that varying the top-squark masses
can influence the radiative BR more drastically in the case of 
the kinematical enhancement (b), due to the presence of considerably 
mixed neutralino states, which increases the relative importance 
of the amplitudes from $t/\tilde{t}$ loops with respect to the ones 
from $\Wpm/\cmp{}$ loops.   
For instance, assuming $m_{\tilde{t}_2} \approx 1$ TeV, increasing 
$m_{\tilde{t}_1}$ from about 100 GeV to about 1 TeV enhances BR(\nrad)
by a few percent, in the case (a). On the contrary, in the case (b), 
the same growth of the lightest top-squark mass gives rise to a wide 
increase (of order 100\%) of the radiative BR. 

\begin{figure}[h]
\centerline{ 
\epsfxsize=0.55\textwidth 
\epsffile{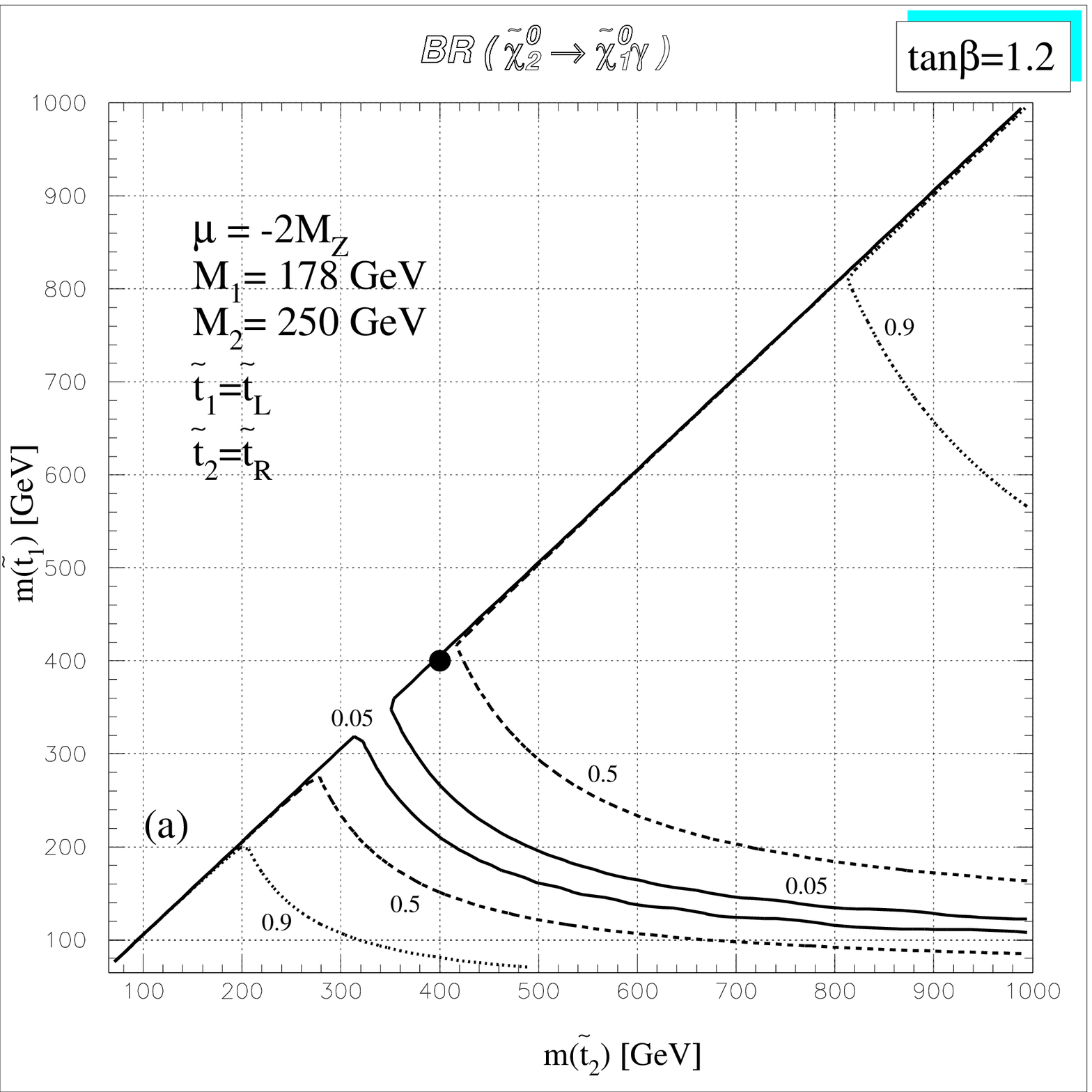}
\hspace{0.5cm} 
\epsfxsize=0.55\textwidth 
\epsffile{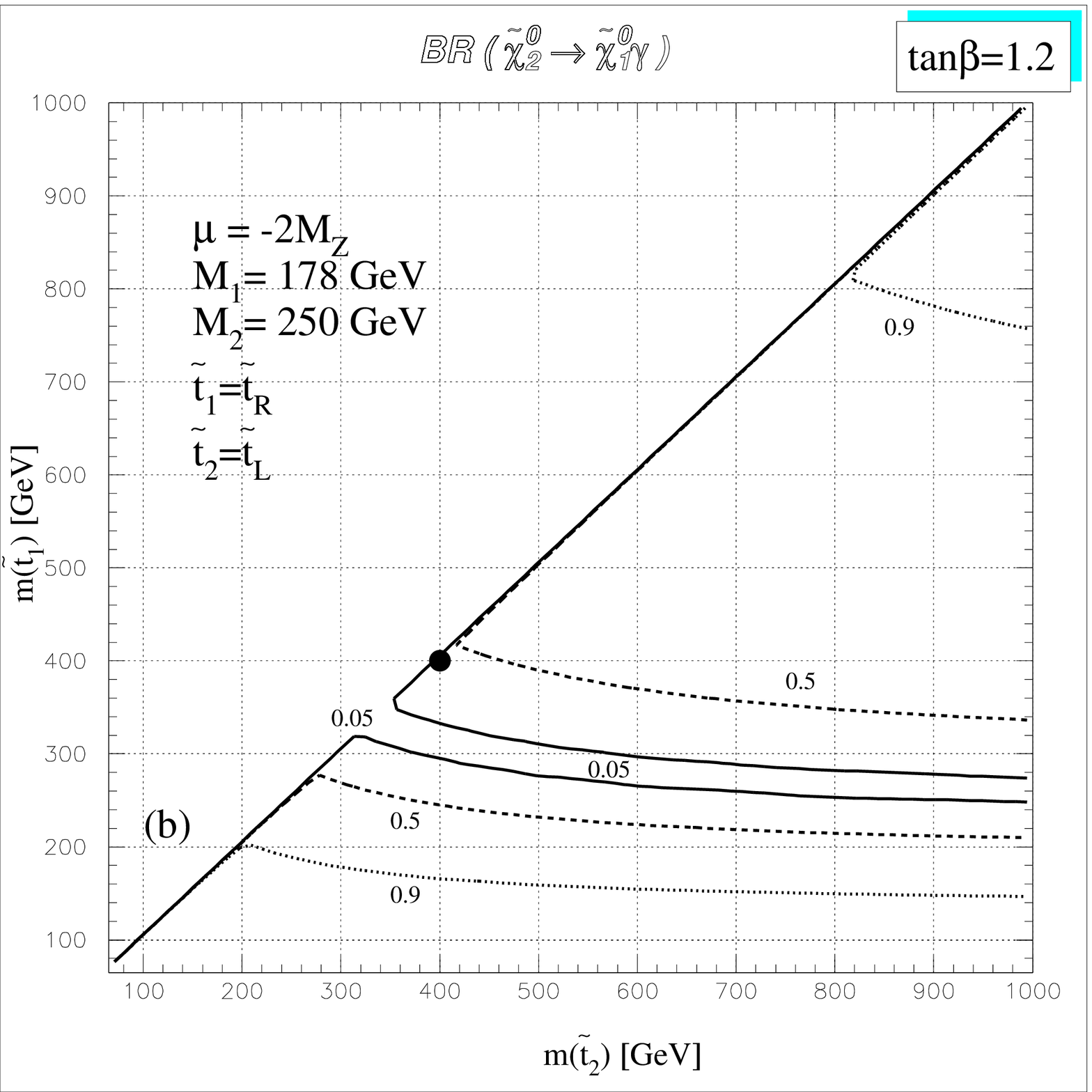}
}
\caption{As in Fig.~12, but for a typical case of kinematical enhancement.  
In (a) the case $\theta_t = 0$, $\protect\tilde{t}_1 = \protect\tilde{t}_L$
is shown; in (b) $\theta_t = \pi/2$, 
$\protect\tilde{t}_1 = \protect\tilde{t}_R$.} 
\label{BRadvstop2}
\end{figure}

Such effect is even more dramatic in the case of Fig.~13. Here, a typical 
scenario with {\it full} kinematical enhancement is shown, with 
$M_1 =178$ GeV, $M_2 = 250$ GeV, $\mu = -2M_Z$, and $\tgb = 1.2$. 
The corresponding, 
nearly degenerate, neutralino masses are $\mn{1} = 180$ GeV and $\mn{2} 
= 185$ GeV. Compared to the cases shown in Fig.~12, in the scenario
of Fig.~13 the relative weight of the $\Wpm/\cmp{}$ and $t/\tilde{t}$ 
loops in the radiative decay matrix element is rather different. 
Indeed, the $\Wpm/\cmp{}$ and $t/\tilde{t}$ amplitudes depend on  
$(\mc{}/M_W)^2$ and $(m_{\tilde{t}}/m_t)^2$, respectively 
(cf. Refs.\cite{Haber-Wyler,Old-Radiative}),
and here one has $\mc{1}$ as large as 196 GeV. 
Thus, the $t/\tilde{t}$ and $\Wpm/\cmp{}$ amplitudes turn out to be 
of the same order of magnitude, with different sign. Then,
one expects that for $m_{\tilde{t}_1} \sim 300$ GeV 
(or possibly more, if the heavier top-squark is also rather light) 
the two contributions tend to cancel each
other, drastically reducing BR(\nrad) at the level of a few percent. 
On the other hand, when 
the top-squarks, and in particular $\tilde{t}_1$, are much lighter (heavier) 
than 300 GeV, the $t/\tilde{t}$ ($\Wpm/\cmp{}$) loops dominate, the
destructive interferences are negligible, and the BR(\nrad) can 
comfortably approach the 100\% level. 
This effect is clearly visible in Fig.~13. 
The influence of a  different top-squark mixing 
can be extracted by comparing Fig.~13(a) and Fig.~13(b), where the {\it pure} 
cases  $\tilde{t}_1 = \tilde{t}_L$ and $\tilde{t}_1 = \tilde{t}_R$,
respectively, are considered. 

\section{Conclusions}

In this paper, we showed that \susy\ scenarios where the radiative mode 
for the next-to-lightest neutralino decay is dominant do exist and can be 
naturally realized, especially when relaxing the electroweak gaugino
mass unification assumptions at the GUT scale. We found that very large
BR's for \nrad are obtained when $\tgb \simeq 1$ and/or $M_1 \simeq M_2$,
with negative $\mu$. 
When $M_1=\frac{5}{3}\tgwq M_2$, it is still possible to achieve a large 
radiative BR, provided $\tgb \simeq 1$ and $\mu < 0$. 
Two different mechanisms, which have different phenomenological implications, 
can be responsible for the radiative BR enhancement. The dynamical mechanism 
may give rise to signatures including a hard photon and missing energy 
at hadron colliders, where $\n{2}$
can be copiously produced either in association with a $\cpm{i}/\n{j}$
or through the decays of heavier sfermions. On the other 
hand, at LEP2, in a scenario with BR(\nrad) dynamical enhancement,
the main neutralino production channel $\epem \to \n{1}\n{2}$
(giving rise to a $\gamma + \miss{E}$ signature) is quite depleted, 
requiring either gaugino or Higgsino sizeable components in both $\n{1}$
and $\n{2}$.  
At larger c.m. energy $\epem$ colliders [like the proposed Next Linear 
Collider (NLC)], the dynamical enhancement of the radiative $\n{2}$ decay 
may be relevant when the $\n{2}$ is produced in association 
with a $\n{j}$ with $j > 1$, giving rise to signatures containing at least a 
hard photon and missing energy. 
On the other hand, the kinematical enhancement (which implies rather 
degenerate light neutralinos) can give rise to sizeable rates 
for final photons of moderate energy in both hadron and 
$\epem$ collisions, provided the next-to-lightest neutralino is produced 
with energy large enough to boost the rather soft photon. 

Finally, an enhanced $\n{2}$ decay into a photon might be of relevance 
for explaining, in the MSSM framework, the \eegg\ event observed by CDF
at the Fermilab TeVatron, that is presently one of the most interesting
hints for possible physics beyond the SM. 

\newpage

\leftline{\bf Acknowledgements}
\vspace{0.3cm} 
\noindent
We would like to thank T.~Gherghetta, G.~L.~Kane, G.~D.~Kribs, and  
S.~P.~Martin for useful discussions and suggestions. 
S.A. was supported mainly by the I.N.F.N., Italy. 
S.A. also thanks Professor Gordy Kane and the Particle Theory Group 
at the University of Michigan for hospitality and additional support.

\end{document}